\documentclass[aps,prx,amsmath,amsfonts,twocolumn]{revtex4-1}%

\usepackage{mathtools}
\usepackage{verbatim}
\usepackage{hyperref}

\usepackage{color}
\usepackage{graphicx}
\usepackage{braket}
\usepackage{amsthm}
\usepackage{amssymb}
\usepackage{amsmath}
\usepackage{ulem}
\usepackage{tikz}
\usepackage{subcaption}
\usepackage{upgreek}

\DeclareMathOperator{\Tr}{Tr}

\newcommand{\old}{\color{black}}

\begin{document}

\title{Measurement Induced Chirality II: Diffusion and Disorder}

\author{Brian J. J. Khor$^{1}$}
\email{bk8wj@virginia.edu}
\author{Matthew Wampler$^{1}$, Gil Refael$^{2,3}$}
\author{Israel Klich$^{1}$}
\email{ik3j@virginia.edu}
\affiliation{$^1$Department of Physics, University of Virginia, Charlottesville, Virginia 22903, USA
}
\affiliation{$^2$Department of Physics, California Institute of Technology, Pasadena CA 91125, USA}
\affiliation{$^3$Institute for Quantum Information and Matter,
California Institute of Technology, Pasadena CA 91125, USA}

\begin{abstract}
   Repeated quantum measurements can generate effective new non-equilibrium dynamics in matter.  Here we combine such a measurement driven system with disorder. In particular, we investigate the diffusive behavior in the system and the effect of various types of disorder on the measurement induced chiral transport protocol \cite{wampler2021stirring}. 
   We begin by characterizing the diffusive behavior produced by the measurements themselves in a clean system. We then examine the edge flow of particles per measurement cycle for three different types of disorder: site dilution, lattice distortion, and disorder in onsite chemical potential. In the quantum Zeno limit, the effective descriptions for the disordered measurement system with lattice distortions and random onsite potential can be modelled as a classical stochastic model, and the overall effect of increasing these disorders induces a crossover from perfect flow to zero transport. On the other hand if  vacancies are present in the lattice the flow of particles per measurement cycle undergoes a percolation phase transition from unity to zero with percolation threshold $p_c \approx 0.26$, with critical exponent $\nu \approx 1.35$. We also present numerical results away from Zeno limit and note that the overall effect of moving away from the Zeno effect is to reduce particle flow per cycle when the measurement frequency in our protocol is reduced.  
\end{abstract}


\maketitle

\section{Introduction}

The development of engineering novel quantum systems by applying periodic (Floquet) driving has produced quantum phases without a static analog \cite{rudner2013anomalous, McIver_2019, Nuske_2020, lindner2011floquet, Titum_2015, kundu2020quantized, nathan2019nathan, po2016chiral,Else2020DTC}. A prominent example is the anomalous Floquet topological insulator \cite{Titum_2015, kundu2020quantized, nathan2019nathan, po2016chiral}, where a chiral edge state emerges alongside completely trivial bulk bands, in stark contrast to standard topological insulators. 
 The idea behind this phase is to break time reversal symmetry by sequentially modulating particle hopping on a lattice; this stirs the particles in such a way that their trajectories in the bulk trace out closed loops, whilst on the edge chiral states emerge \cite{rudner2013anomalous}. 
 Such dynamics has been realized experimentally in, for example, cold atom systems \cite{Wintersperger2020ExperFloq,braun2023ExperimentFloq}, while theoretically these ideas have recently been extended to interacting systems where an even more diverse class of topological phases emerges \cite{nathan2019nathan,Nathan2021AFI,wampler2022arrested, wampler2022fragmentations}.

On the other hand, the interplay between measurements and unitary time evolution in quantum many-body systems has received renewed interest in recent years \cite{ wampler2021stirring, agrawal2022monitoredcircuit, Biella2021manybodyquantumzeno, block2022measurementlongrange, chan2019unitaryprojective, doggen2022measurementMPS, ippoliti2021measurementonly, jian2020measurementcriticality, ladewig2022monitoredopenfermion, li2018quantumzeno, lu2021localizedmeasurement, lunt2021measurementpercolation, minato2022measurementlongrange, minoguchi2022measurementbosoncft, shtanko2020classicalmonitored, skinner2019measurement, szyniszewski2019entanglement, Sierant2022dissipativefloquet, skinner2019measurement, turkeshi2020measurementcriticality2d,  zabalo2020measurement, fisher2022review,alberton2021moniteredfermionchain, buchhold2021theorymeasurementfermion, cao2019fermionmonitoring, chen2020conformal, Fidkowski2021howdynamicalquantum, kells2021topologicalmonitoredfermion, szyniszewski2022disorderedmonitor, merritt2022freefermions, yang2022nlsmmeasurement}.  This is, in part, due to developments on phase transitions in the entanglement entropy of random unitary circuits with measurements (see \cite{fisher2022review} and references therein) as well as on the utility of measurements to induce non-trivial dynamics and to prepare quantum states \cite{wampler2021stirring,popperl2022measurementAnderson,yamaguchi2023StatePrep,Garratt2023MeasDyn,Riste2013StatePrep,Roy2020StatePRep}.

In \cite{wampler2021stirring}, it was shown that periodic sequences of measurements may be used to induce chiral edge charge transport alongside trivial bulk dynamics in a way much analogous to anomalous Floquet insulators.  The general intuition behind this procedure is to use measurements to control the effective particle hopping on a lattice.  This may be seen most clearly in the limit of rapid measurements, the quantum Zeno limit, where dynamics is frozen within monitored sections of the lattice and hopping is eliminated between monitored and un-monitored sections of the lattice.  This measurement-based control of the particle hopping may then be leveraged to recreate the periodic modulation of hopping amplitudes used to induce anomalous Floquet insulators.  However, the measurement-based scheme also exhibits distinct features due to the non-unitary nature of the evolution.       

In this work, we continue an investigation of the measurement-induced chirality protocol \cite{wampler2021stirring} by looking at the following aspects.  (1) The diffusive dynamics of the measurement-induced chiral systems when the system is tuned away from the 'perfect swapping' limit and away from the Zeno measurement limit, and (2) the effects of various kinds of disorder on the measurement-induced chiral flow rate in our free fermion systems hopping on a Lieb lattice. In particular, we consider chiral flow in the case of site vacancy disorder, random hopping strength, and random onsite potential, both in and out of the Zeno measurement limit, and diffusive dynamics for the case of random onsite potential. Indeed, an important characterization of systems exhibiting chiral physics is their response to disorder, as have been long studied in, e.g. the context of the quantum Hall effect \cite{trugman1983localization,Chalker_1988}, where relations to percolation physics have been explored. Disorder also plays a crucial role in the anomalous Floquet insulators mentioned above.

\begin{figure}
    \centering
    \includegraphics[width=0.49\textwidth]{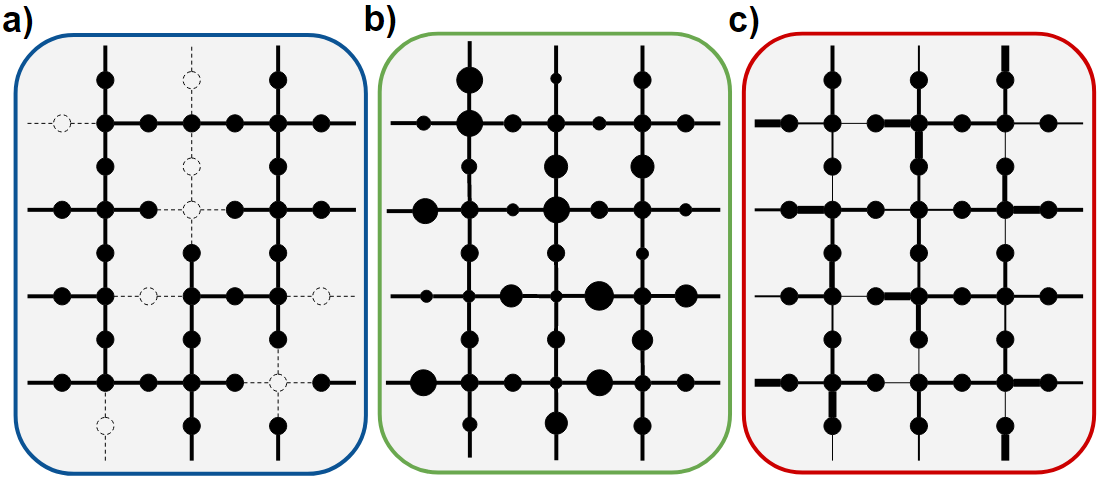}
    \caption{The three varieties of disorder considered in this work: a) Sites are randomly removed from the lattice with probability $p$.  b) Random on-site potentials are applied to the lattice (the strength of the potential at each site is represented by the size of the vertex) c) The hopping strength between adjacent sites is given by a uniform, random distribution (represented by the size of the edge).}
    \label{fig: disorder summary}
\end{figure}

The approach we take in this paper will import techniques developed in \cite{KLICH201966}, in similar spirit taken in our earlier work \cite{wampler2021stirring}. In \cite{KLICH201966}, the competing effects of unitary evolution and measurements were studied using a closed hierarchy approach. This technique has also been used, for example, to describe non-equilibrium steady states of current \cite{KLICH201966} and density fluctuations (quantum wakes) following a moving particle detector and other disturbances \cite{wampler2020wakes}.

The structure of our work is as follows. In section II, we briefly review the measurement protocols and the basic physics behind our earlier work on measurement-induced chirality \cite{wampler2021stirring}. This is followed by studying the diffusive dynamics of the measurement-induced chirality out of the 'perfect swapping' parameter and out of Zeno limit in section III. After dealing with clean systems, we proceed to study 2 variesties of disorders in subsequent sections as illustrated in Fig \ref{fig: disorder summary}. In section IV, we deal with site vacancy disorder for our system in the Zeno limit, as motivated by our system. In particular, there is a percolation threshold when measurement period is tuned to 'perfect swapping' case with deterministic walk \cite{PhysRevE.104.064122}. In section V, we numerically simulate the effect of random hopping strength and random onsite potential on the chiral flow rate induced by measurement, still operating in the Zeno limit, and provided an analytical mean field treatment to describe the weak disorder limit. In section VI, we investigate numerically the diffusive dynamics and the chiral flow rate for our measurement-induced chiral system under all three types of disorder (vacancy, random hopping and random potential) away from the Zeno limit. We present discussions and possible outlook in section VII.

\section{Measurement Induced Chirality Protocol}
\label{Sec: Review of Stirring}

In this section, we briefly review the protocol realizing the measurement induced chirality in \cite{wampler2021stirring}. We consider fermions freely hopping on a Lieb lattice, subject to a cycle of local density measurements as follows.  

\begin{figure}[h]
    \centering
    \includegraphics[width=0.5\textwidth]{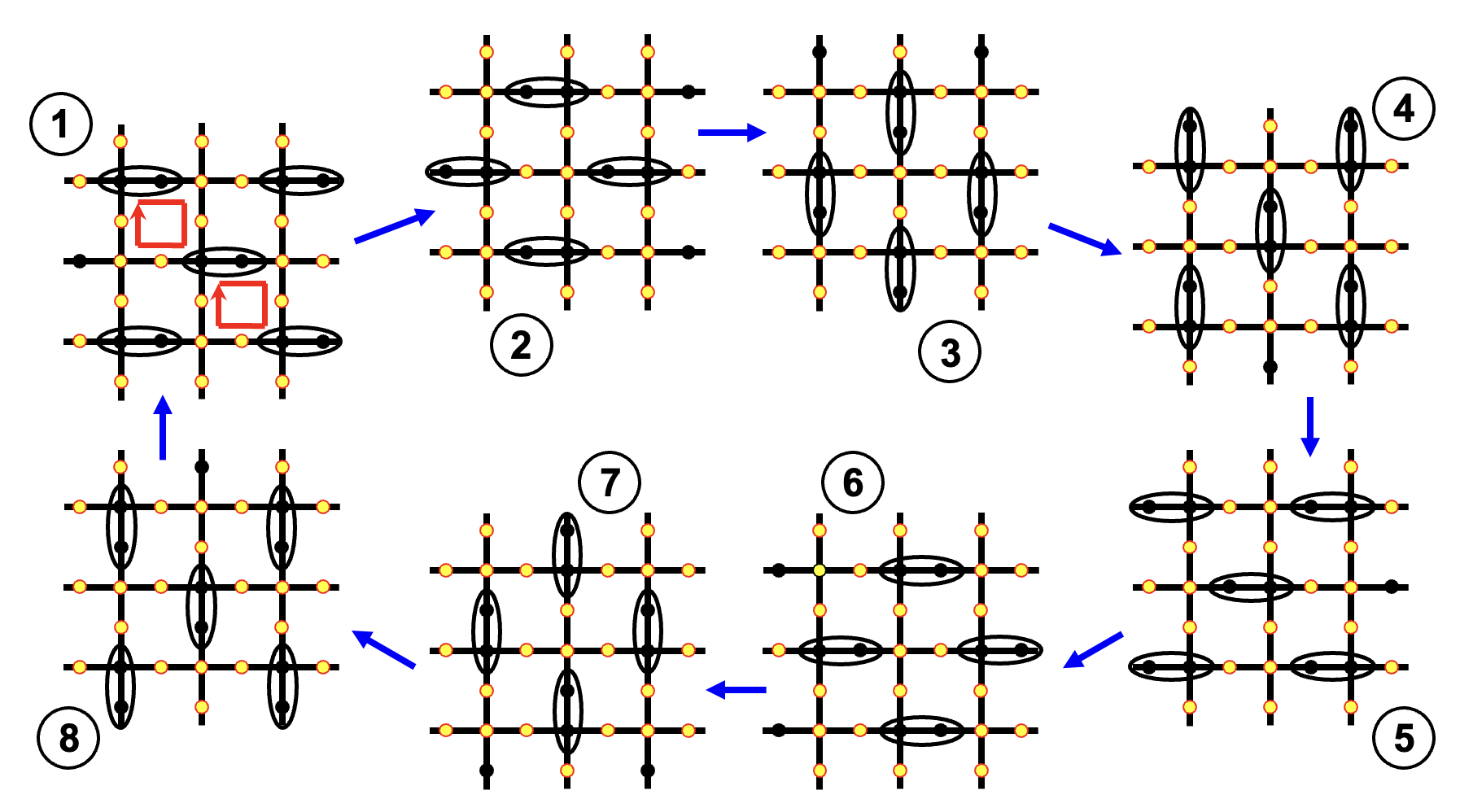}
    \caption{Measurement protocol. Yellow vertices indicate the set of repeatedly measured sites, while black sites are the unmeasured, free evolving set, $A_i$. The adjacent black vertices trace out a chiral path around a plaquette in the Lieb lattice. In the Zeno limit with perfect swapping parameters, particles will trace out the path as shown by the red loops in first figure.}
    \label{fig: meas_protocol}
\end{figure}

The measurement cycle in Fig \ref{fig: meas_protocol} consists of 8 steps taking an overall measurement period $T$. At each step, we take repeated measurements to detect particles throughout a subset of the lattice, while the system is allowed to evolve freely in between measurements with the nearest neighbour hopping Hamiltonian ${\cal H} = - t_{hop} \sum_{\langle i, j \rangle} c_i^{\dagger} c_j + \text{h.c.}$. We denote the set of sites {\it not} being measured at step $i$ by $A_i$ as marked in Fig (\ref{fig: meas_protocol}) and enforce periodicity by setting $A_{i+8} = A_i$.  Within step $i$, we carry out the following steps:  
\begin{enumerate}
    \item Particle densities at all sites in $(A_i \cap A_{i-1})^c$ are measured, i.e., we measure all other sites in the lattice \textit{except} the sites circled in each step in Fig \ref{fig: meas_protocol} 
    \item Free evolution under a free hopping Hamiltonian ${\cal H} = -t_{hop} \sum_{\langle {\mathbf{r}}{\mathbf{r'}} \rangle } c^{\dagger}_{\mathbf{r}} c_{\mathbf{r'}} + \text{h.c.}$ for a time $\tau=\frac{T }{8 n}$. Here $n$ is an integer describing the measurement frequency. 
    \item Particle densities at all sites in $A^c_i$ are measured.
    \item Steps 2 and 3 are repeated $n$ times. 
\end{enumerate}

The overall effect of the measurement protocol has been shown in \cite{wampler2021stirring} to exhibit protected chiral charge flow of the particles,  as shown by the red loops in Fig \ref{fig: meas_protocol}. The physical intuition of measuring everywhere in the lattice other than the circled sites is to restrict (and in the case of rapid Zeno monitoring, freeze entirely) particle motion elsewhere other than between the circled pairs of sites. This dynamics is reminiscent of the Anomalous Floquet Topological Insulator \cite{rudner2013anomalous}   in that we selectively switch on certain links on the lattice where free evolution are allowed to take place, but also presents differences due to the non-unitary nature of the measurements. 

We emphasize that the Lieb lattice was chosen for ease of comparison with the Floquet insulator dynamics of \cite{rudner2013anomalous}, however the particularities of the band structure associated with the Lieb lattice (a flat band) are not important since any coherent dynamics is quickly disrupted by measurement as we will see in detail.    It is only for reduced measurement rates that effects from the band structure may begin to emerge.\old We also remark that the measurement protocol is not restricted to the Lieb lattice alone. For a more thorough discussion on the geometric conditions a lattice needs to satisfy in order to carry out a measurement protocol as above, we refer interested readers to Appendix C in \cite{wampler2021stirring}.

In order to study the charge flow, we focus on the dynamics of the two point correlation function $G(t)_{\mathbf{r} \mathbf{r'}} = \Tr(\rho (t) c_{\mathbf{r}}^{\dagger} c_{\mathbf{r'}})$ under measurement and unitary time evolution. The correlation $G$ transforms in a simple way under particle detection measurements and under  non-interacting evolution, respectively (see e.g. \cite{KLICH201966, wampler2021stirring}):
\begin{eqnarray}
    &G \rightarrow (1-P_{{\mathbf{r}}}) G (1-P_{{\mathbf{r}}}) + P_{{\mathbf{r}}} G P_{{\mathbf{r}}} , \label{eq: meas G}\\
    &G \rightarrow  U G U^{\dagger} ,\label{eq:G unitary}
\end{eqnarray}
where $P_{\mathbf{r}}=\ket{\mathbf{r}}\bra{\mathbf{r}}$ is the projector onto site $\mathbf{r}$ where a particle detection measurement is has been performed,  and $U = \exp(- i H t)$ is the (single particle) unitary time evolution between consecutive measurements under the free Hamiltonian $H$ for time $t$. We stress that the map \eqref{eq: meas G} is the result of averaging over measurement outcomes.

It is informative to consider the limit of many measurements per step ($n\rightarrow \infty$), i.e. the quantum Zeno limit. The signature characteristic of this regime is the freezing of evolution in the subspace of measured sites while free evolution continues to occur between unmeasured sites.  In other words, the time evolution during the $i$-th step of the 8 step measurement cycle may be replaced by evolution under the Hamiltonian
\begin{eqnarray}
    &{\cal H}_{A_i} = -t_{hop} \sum_{\langle {\mathbf{r}}{\mathbf{r'}} \rangle \in A_i } c^{\dagger}_{\mathbf{r}} c_{\mathbf{r'}} + \text{h.c.} ,
\end{eqnarray}
where now the evolution is confined to between unmeasured sites within each set $A_i$. 

Another important aspect of the dynamics in the Zeno limit is that in this limit, repeated application of Eq. \eqref{eq: meas G} under our protocol, kills off-diagonal elements of $G$ between sites in the set ${A_i}^{c}$, and the  switch between measurement of ${A_i}^{c}$ to the next step $A_{i+1}^{c}$ eliminates any lingering  off-diagonal correlations that have developed inside the set $A_i$ during the evolution \cite{wampler2021stirring}.  Therefore, at the beginning/end of steps we  only need to keep track of the diagonal components $G_{{\bf rr}}$. Let us combine these in a vector, $\ket{g(t)}$, where $\braket{{\bf r}|g(t)}\equiv  G_{{\bf rr}}$. Then the effective action of step $i$ in the protocol, including the unitary evolution and measurements in the Zeno limit, is described by 
\begin{align}   &
\ket{g(t)}\rightarrow R \ket{g(t)} \\ &
    R_i=\oplus_{\langle\mathbf{r},\mathbf{r'}\rangle\in  A_i}\left(
    \begin{array}{cc}\small
        1-p_{hop}  &  p_{hop}  \\
        p_{hop}  & 1-p_{hop} \\
    \end{array}
    \right) \oplus_{\text{other sites}} I ,
    \label{eq: stochastic}
\end{align}
i.e. the evolution of the local particle density $\ket{g_i (t)}$ in the Zeno case is given by a periodically driven random walk.  The probability for hopping between sites is related to the period of each measurement step $T/8$ by the following equation \cite{wampler2021stirring}
\begin{eqnarray}
    p_{hop} = \sin^2(\frac{T}{8}) . 
    \label{eq: hopping prob}
\end{eqnarray}   
Note that when the full measurement period $T=4\pi (2n+1), n\in\mathbb{Z}$, we have $p_{hop}=1$ and the evolution becomes deterministic hopping/walk, a situation we call "perfect swapping". Similarly, when $T=8 \pi n$ the evolution is frozen, with $p_{hop}=0$. We now summarize the two methods used in \cite{wampler2021stirring} to measure the chiral charge transport induced by the measurement protocol in the system. 

\begin{figure}[h]
    \centering
    \includegraphics[width=0.4\textwidth]{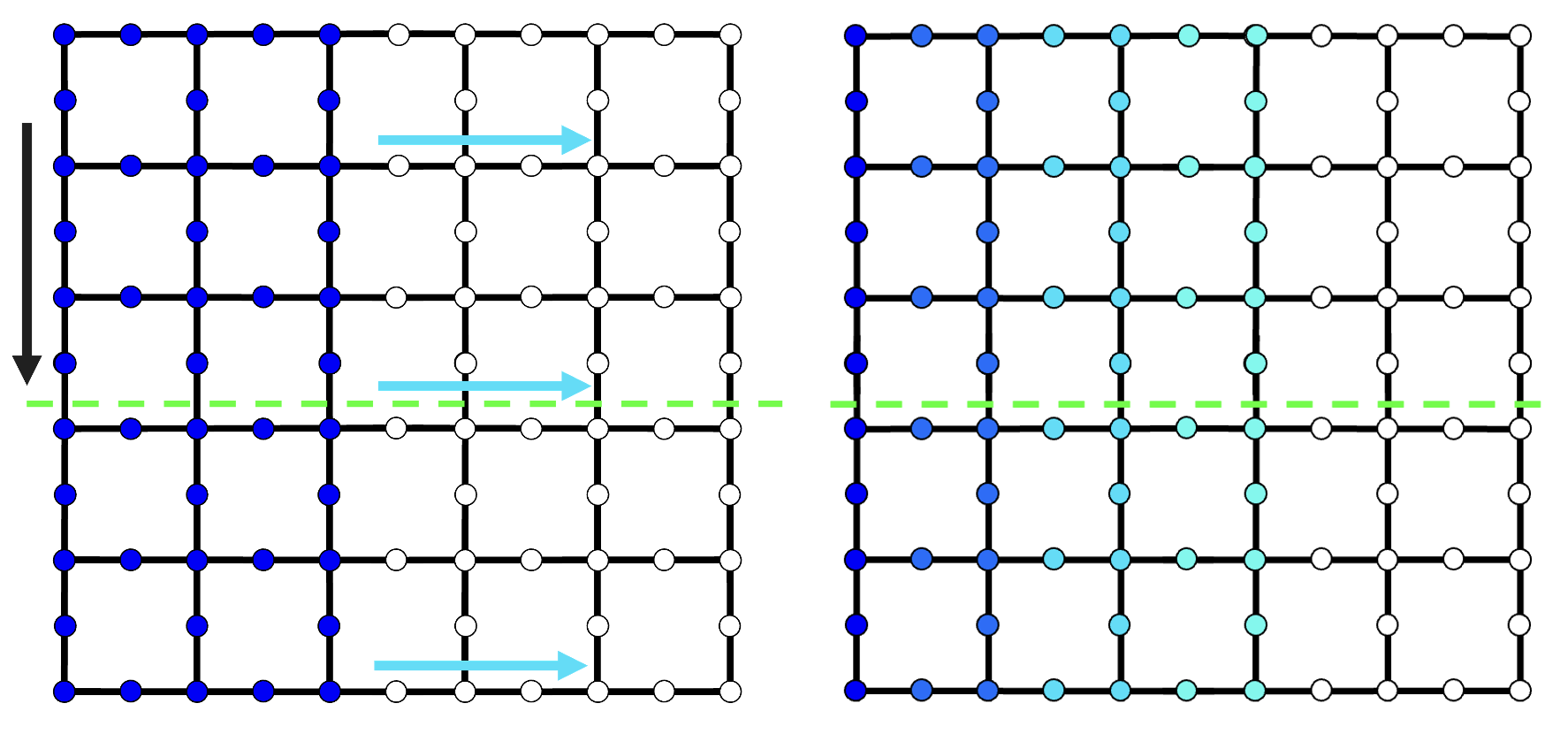}
    \caption{Left: The initial configuration and setup for studying chiral particle flow for all disorder cases considered (with left half plane filled with particles in blue). The particle exhibits both downward chiral motion (black arrow) and diffusive motion which moves the front to the right (blue arrows). Right: After running the protocol in the half-filled lattice setting, particles density increases beyond the initial configuration, where lighter blue indicates lower particle density.} 
    \label{fig: Flow Diagram}
\end{figure}

In the first method, the charge flow is found numerically by making a cut through the lattice and measuring the charge flow across it, Fig. \ref{fig: Flow Diagram}.  Namely, the number of particles flowing across the slice is found by measuring the change in total particle number below the cut, i.e. given by
\begin{equation}
    F_{sim}(t) \equiv \sum_{\mathbf{r} \text{ below slice}} \big( G(t)_{\mathbf{r r}}- G(t=0)_{\mathbf{ r r}} \big) .
    \label{eq: Fsim}
\end{equation}
In order to measure the charge flow along a given edge in the system, we fill up all the sites near that edge with particles and then measure \eqref{eq: Fsim} while applying the measurement protocol.  The filling of half the system with particles must be done since, if the whole system was filled, opposite edges in the system would exhibit charge transport with equal magnitude but opposite direction, thereby leading to a no net flow of particles across the cut.  The details of how particles are inserted into the bulk of the system and the specific path of the cut through the lattice do not affect the charge flow per measurement cycle (beyond transient effects) \cite{wampler2021stirring}.  

The second method used to measure the chiral particle flow assume rapid measurements and relies on the counting statistics of the transport up or down in the system. We introduce a counting field $e^{i\theta}$ to each vertical link by modifying the transition matrices $R_3,R_4,R_7,R_8$ to

\begin{eqnarray}  
 R_{i }=\oplus_{\langle \mathbf{r}, \mathbf{r'}\rangle\in  A_i}\left(
\begin{array}{cc}
 1-p_{hop}  & e^{i\theta}p_{hop}  \\
 e^{-i\theta}p_{hop}  & 1-p_{hop} \\
\end{array}
\right)\oplus_{\text{other sites}} I   \label{eq:Ri with theta} 
\end{eqnarray}
whenever $\mathbf{r}, \mathbf{r'}$ are nearest neighbours on a vertical line such that site $\mathbf{r}$ is located above $\mathbf{r'}$.  We will denote the transition matrix (with counting fields) of the full measurement cycle by

\begin{gather}
\label{eq: Rcyc}
    R_{\text{cyc}}(\theta) = R_8 R_7 R_6 R_5 R_4 R_3 R_2 R_1
\end{gather}
   
With the counting field present, we can introduce the moment generating function after $N$ measurement cycles 
\begin{eqnarray}&
\chi_{N}(\theta) =  \sum_{\mathbf{r} \mathbf{r'}} [R_{cyc}(\theta)^N]_{\mathbf{r} \mathbf{r'}}   G_{\mathbf{r} \mathbf{r}}(t=0)
\end{eqnarray}
which may be used to calculate the charge transport in the $y$ direction.  Namely, the flow per unit length per measurement cycle (in the long time, $N \rightarrow \infty$, limit) is given by 
\begin{eqnarray}
   F = \lim_{N\rightarrow \infty} {1\over L_y}{1\over N }\left. i \partial _{\theta }  \chi_N(\theta) \right|_{\theta=0} \label{eq: F equation}
\end{eqnarray}
with $L_y$ the length of the system in the $y$ direction.

In \cite{wampler2021stirring}, it was shown that the analytical form of the flow in the Zeno limit exhibits bulk-edge decomposition in the sense that $F$ can be decomposed into a term that is calculated entirely with bulk operators only and another term that is computed from the edge operators alone
\begin{equation}
    F = F_{bulk} + F_{edge} .
    \label{eqn: F_analytic}
\end{equation}
A computation performed in \cite{wampler2021stirring} shows the flow $F_{edge}$ depends on $p_{hop}$ via
\begin{equation}
    F_{edge} = p_{hop}^2 + p_{hop}^3 + p_{hop}^4 .
\end{equation}
The other flow term $F_{bulk}$, on the other hand, depends on $p_{hop}$ in a more nontrivial way, and it is best instead to express $F_{bulk}$ in the following form in terms of the bulk operators
\begin{equation}
    F_{bulk} = i \sum_{\alpha \beta} \left. \left[ J_B (\mathbf{k}) \frac{1}{I - R_B (\mathbf{k})} \partial_{k_y} R_B (\mathbf{k}) \right]_{\alpha \beta} \right|_{\mathbf{k}=0} .\label{eq:bulk contribution}
\end{equation}
Here, $R_B (\mathbf{k}) = R_B (\mathbf{k}, \theta = 0)$ is a bulk transition operator, equivalent to $R_{cyc} (\theta)$ in Eqn (\ref{eq: Rcyc}) except that it comes with periodic boundary conditions instead of open boundary conditions. In equation \eqref{eq:bulk contribution} $R_B$ is expressed in k-space. The explicit construction of $R_B (\mathbf{k}, \theta)$ is delineated in Appendix \ref{Appendix: Diffusion} as a $6 \times 6$  matrix in terms of $p_{hop}$, $\mathbf{k}$ and the counting field $\theta$. Here, $J_B (\mathbf{k}) = - i \partial_{\theta} R_B (\mathbf{k}, \theta) |_{\theta = 0}$. From the expressions of $J_B (\mathbf{k})$ and $R_B (\mathbf{k})$ (as a nontrivial matrix of $p_{hop}$) one can then compute $F_{bulk}$ from a given $p_{hop}$ as a sum of the resulting matrix elements. This formalism is summarized rather briefly here and we refer readers to \cite{wampler2021stirring} for a more extensive discussion and proof.

\section{Measurement induced diffusion}


In \cite{wampler2021stirring} (as reviewed in Sec. \ref{Sec: Review of Stirring}), the focus was on the emergence of the protected, chiral transport near the edge of the system.  However, it is worth analyzing further the dynamics in the bulk.  

One reason is that the bulk dynamics sets the time scale over which the chiral edge transport is sustained.  To see this, take for example the initial particle configuration described in Fig. \ref{fig: Flow Diagram}. Note that if any of the holes initially located in the right half plane of the lattice reach the left boundary of the system at some time during the evolution, then the hole may be transported along the edge in place of a particle.  This would then alter (namely, reduce) the edge flow.  Hence, the time scale over which the flow $F$ is robust is set by the length of time it takes for holes initially in the bulk to reach the boundary.  In this section, we study the diffusive behavior of bulk particles (and holes).  We calculate the diffusion coefficient  analytically in the Zeno limit of our measurement protocol and find it numerically for finite measurement frequencies. 

\begin{figure}[h]
    \centering
    \includegraphics[width=0.48\textwidth]{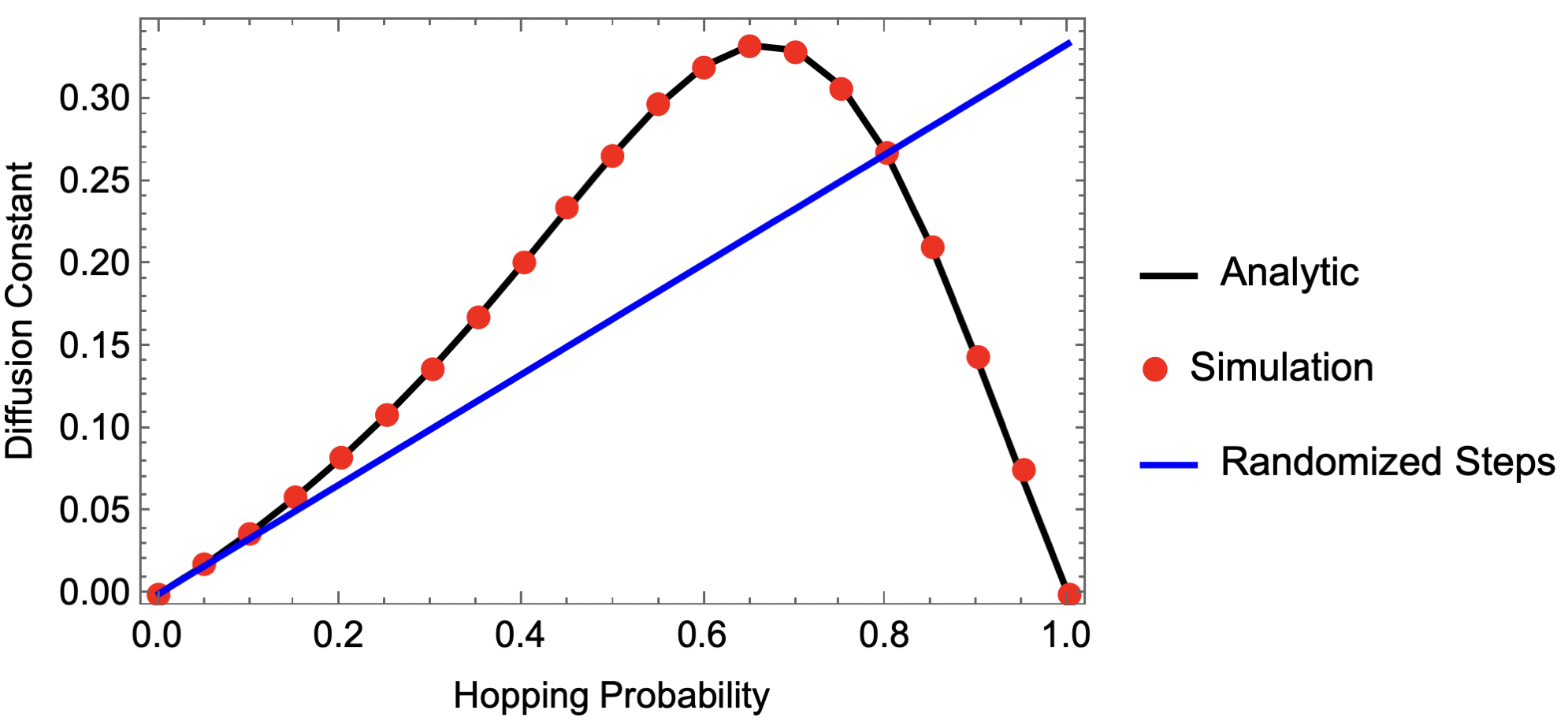}
    \caption{The diffusion constant as a function of hopping probability for measurement induced chirality in clean system, in the Zeno limit. The averaged randomized quantum Zeno measurement protocol is shown in the blue line.} 
    \label{fig: Diffusion Constant}
\end{figure}

Another reason to take a closer look at the dynamics in the bulk, is that it acts as a further probe of the interplay between the chirality of the measurement scheme and the stochastic behavior induced from the random measurement outcomes. The juxtaposition of these two effects was particularly clear when analyzing the edge dynamics.  For example, working in the Zeno limit, consider replacing the chiral stochastic evolution $R_{cyc}$ \eqref{eq: Rcyc}, by 
evolving with a randomized protocol,  where in each step of the 8-steps protocol we randomly pick $R_i$ and average over all possible outcomes. This situation is described by the averaged cycle $\overline{R_{cyc}} = \frac{R_1 + R_2 + R_3 + R_4 + R_5 + R_6 + R_7 + R_8}{8}$. In this case, the random walk exhibits diffusive dynamics even in the perfect swapping case, in sharp contrast with the chiral protocol  Fig \ref{fig: Diffusion Constant}.  It is truly the chiral nature of the drive that then is responsible for the ballistic transport along the edge.  In the bulk of the system, transport is diffusive as it would be without the introduction of chirality into the drive, but nonetheless the chirality does still play a role leading to different diffusion constants. \old


\begin{figure}[h]
    \centering
    \includegraphics[width=0.45\textwidth]{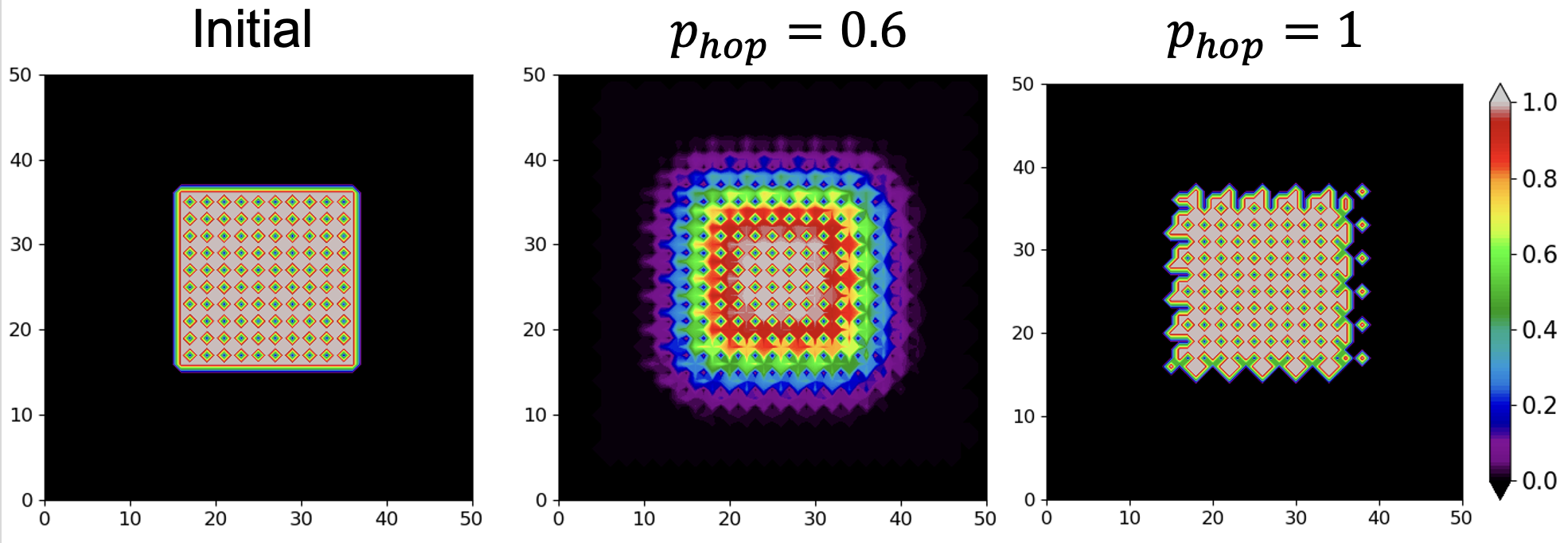}
    \caption{A finite droplet is allowed to evolve under quantum Zeno measurement and unitary evolution. The droplet configuration for $p_{hop} = 0.6$ exhibits both outward diffusion and clockwise chiral edge transport while the droplet for $p_{hop} = 1$ exhibits clockwise chiral edge motion only. The droplet setup is employed for all calculations and simulations on the diffusion constant.} 
    \label{fig: Diffusion Image}
\end{figure}

A clarifying comment is in order about different origins of the diffusive dynamics in the system. Namely, there are two distinct ways to induce diffusion discussed in this paper: through the measurements (via the tuning of the length of each measurement step away from perfect swapping or by reducing measurement frequency) which occurs already in the clean system, and through the addition of disorder (e.g. on-site potential or hopping strength disorder).  In this Section, we focus on the former, leaving a discussion of disorder-induced diffusion to Sections \ref{Sec: lattice distortion and on-site disorder} and \ref{Sec: Disorder Away From Zeno}. \old    

\begin{figure}[h]
    \centering
    \includegraphics[width=0.46\textwidth]{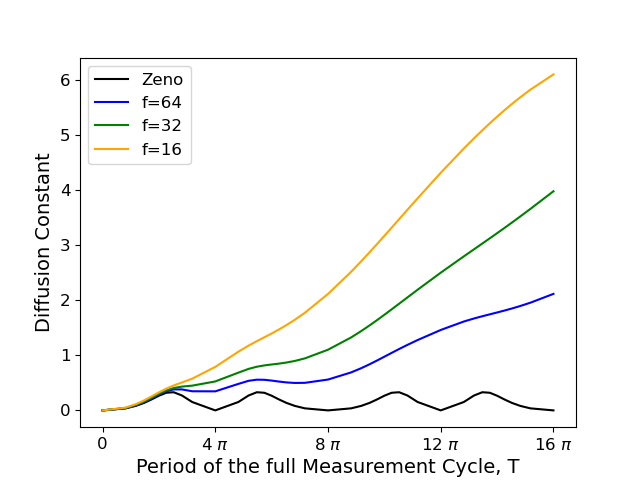}
    \caption{The diffusion constant in a measurement induced chiral Lieb lattice as a function of the period of measurement cycle $T$ up until $T = 16 \pi$, where in the Zeno limit the diffusion constant is periodic with $8 \pi$. Various measurement frequencies per measurement step, $f$, is shown.} 
    \label{fig: Diffusion Constant Non Zeno}
\end{figure}

Transport in a system is defined as diffusive when the average squared displacement from the center of mass is linear in time, i.e. 
\begin{gather}
    \langle \Delta r^2 (t) \rangle = Dt ,
\end{gather}
where $D$ is the diffusion constant. In our case, starting with $G_{\mathbf{r} \mathbf{r}}(t=0)=\delta_{\mathbf{r} \mathbf{r'}}\delta_{\mathbf{r},0}$, we have
\begin{eqnarray}
    &\langle \Delta r^2 (t=NT) \rangle \equiv \frac{\sum_{\mathbf{r}} ( \mathbf{r} - \mathbf{r}_{mean} (t=NT) )^2 G_{\mathbf{r} \mathbf{r}} (t=NT) }{\sum_{\mathbf{r}} G_{\mathbf{r} \mathbf{r}} (t=NT)} , \nonumber \\ 
    &\mathbf{r}_{mean} (t=NT) \equiv \frac{\sum_{\mathbf{r}} \mathbf{r} G_{\mathbf{r} \mathbf{r}} (t=NT) }{\sum_{\mathbf{r}} G_{\mathbf{r} \mathbf{r}} (t=NT)} , \nonumber \\
    &D = \lim_{N \rightarrow \infty} \frac{\langle \Delta r^2 (t=NT) \rangle - \langle \Delta r^2 (t=0) \rangle }{NT} . \label{eq: diffusion def}
\end{eqnarray}
Numerical results for the diffusion constant are shown in Fig. \ref{fig: Diffusion Constant Non Zeno}, for both Zeno limit and finite measurement frequencies. 

As one reduces the measurement frequency per measurement step away from the Zeno limit, the diffusion constant increases. The feature where diffusion transport is suppressed (or absent in the Zeno limit) at $T = 4 \pi n, n \in \mathbf{Z}_{+}$ becomes less pronounced in low frequency limit, and eventually the diffusive transport will become ballistic without any measurement. 

In the Zeno limit, the absence of diffusion at $T = 4 \pi n, n \in \mathbf{Z}_+$ can be attributed either to the perfect swapping case where $n \in \text{odd}$, or the zero hopping case where $n \in \text{even}$ by inspecting Eqn \ref{eq: hopping prob}, with $p_{hop} = \sin^2( \frac{T}{8})$ (setting $t_{hop} = 1$). We note that the Zeno limit diffusion constant curve exhibits $8 \pi$ periodicity. 
 
We can extract the diffusion constant both analytically and numerically in the Zeno limit. Numerically, we calculate the diffusion constant on a finite lattice using Eqn \ref{eq: diffusion def} for intermediate time scale before the particle distribution hits the boundary by having nonzero $G_{\mathbf{r} \mathbf{r}}(t) > 0$ for boundary sites $\mathbf{r}$. Analytically, we consider a translationally invariant system with setup considered in Appendix \ref{Appendix: Diffusion}, placing a single particle at the position 1 of the unit cell at origin and performing the calculation in momentum space with the formula (see derivation in Appendix \ref{Appendix: Diffusion})
\begin{eqnarray}
    D &=& \lim_{N \rightarrow \infty} \frac{1}{8N} \left[ \sum_{\mu=1}^6 [ - \nabla^2_{k} R_{B}^N (\mathbf{k}) |_{\mathbf{k} = 0} ]_{\mu, 1} \right. \nonumber \\ 
    &+& \left[ \sum_{\mu=1}^6 [ \partial_{k_x} R_B^N (\mathbf{k}) |_{\mathbf{k}=0} ]_{\mu,1} \right]^2 \nonumber \\
    &+& \left. \left[ \sum_{\mu=1}^6 [ \partial_{k_y} R_B^N (\mathbf{k}) |_{\mathbf{k}=0} ]_{\mu,1} \right]^2 \right]
    \label{eq: diffusion k space}
\end{eqnarray}
where $R_B = R_8 R_7 R_6 R_5 R_4 R_3 R_2 R_1$ is the $6 \times 6$ transition matrix in Eq \ref{eq: stochastic} written in $k$-space (see Appendix \ref{Appendix: Diffusion} for explicit form of $R_i (\mathbf{k})$) by making use of translational invariance. We see that the analytical and numerical result agree well as shown in Figure \ref{fig: Diffusion Constant}. 

Identifying the diffusion coefficient, is also helpful for the numerical calculation of the edge flow we discuss in the next section. Indeed, to correctly extract the late time dynamics of the measurement induced chiral flow, one wants to estimate the maximum time scale $t_{max}$ for which the chiral flow counts mostly only the flow travelling to the lower half of the Lieb lattice setup in Figure \ref{fig: Flow Diagram} before the transverse spreading from diffusion hits the right boundary of the Lieb lattice in Figure \ref{fig: Flow Diagram}. This is done to exclude  finite size lattice boundary effects on the numerical counting of the chiral flow. By taking into account the transverse diffusive transport, we can extract the late time chiral transport for different system sizes while taking into account finite size effect systematically. We discuss how we extract the late time mean flow per cycle for the rest of the paper in appendix \ref{Appendix: Finite Size Late Time}.  

\section{Site dilution and Percolation Threshold}
\label{sec: site vacancy}

\begin{figure*}[t]
    \centering
    \includegraphics[width=1.\textwidth]{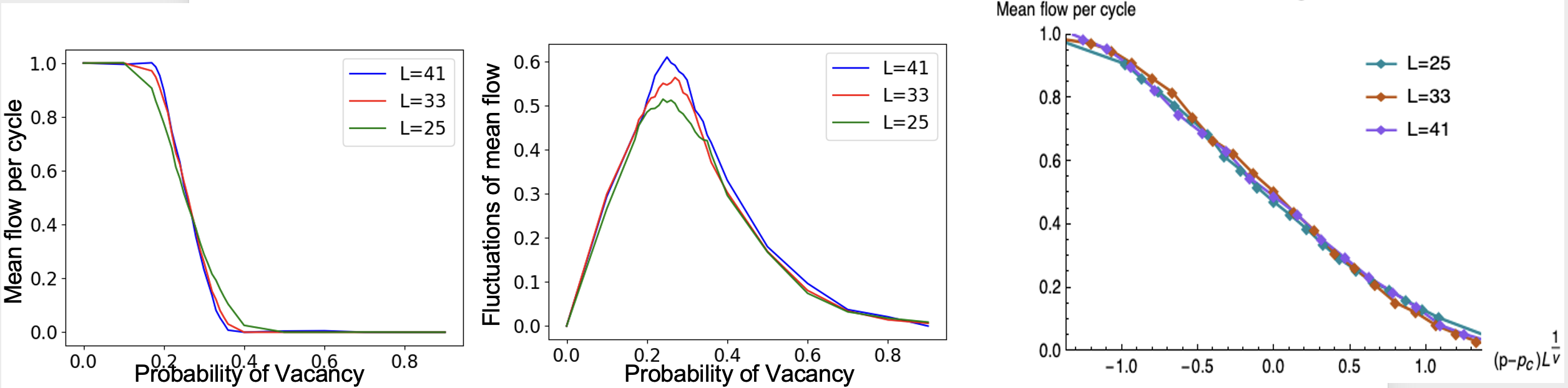}
    \caption{Left: Mean flow per cycle in the perfect swapping case $p_{hop} = 1$ plotted as a function of probability of site vacancy in the lattice $p_{\alpha}$ for various linear system sizes $L_x$ (see Fig \ref{fig: Flow Diagram}) over 1000 disorder realizations. Middle: The fluctuations of the mean flow per cycle across 1000 disorder realizations for various system sizes with peak around $p_c \approx 0.26$. Right: The scaling collapse with the functional form $\langle F \rangle = f( (p_{\alpha}-p_c) L^{1/ \nu} )$, where $p_c \approx 0.26$ and $\nu \approx 1.35$.}
    \label{fig: perfect swapping percolation}
\end{figure*}

In \cite{wampler2021stirring}, the measurement induced chiral charge transport along the edge of the system was shown to be protected against edge perturbations in analogy with the protected edge flow induced in anomalous Floquet insulators.   In order to investigate further the nature of the measurement-induced protection, \old we now turn to consider the effects of several different varieties of disorder on the chiral flow.  Furthermore, these considerations of disorder will also be useful for gaining insight into the effects of imperfections that may occur in real experimental implementations of the measurement protocol. 

The first kind of disorder we consider is site vacancy disorder.  Namely, as shown in Fig. \ref{fig: disorder summary}a, we  \old consider a situation where there is a probability $p_{\alpha}$ for each site on the Lieb lattice to be vacant, i.e. where particles are prohibited from hopping to or from the vacant sites. The locations of the vacant sites then stays constant throughout the measurement protocol.  


In this section, we will only consider the quantum Zeno limit and postpone results away from the Zeno limit until section \ref{Sec: Disorder Away From Zeno}.  
In this case, there are 2 different scenarios: (1) the perfect swapping case ($p_{hop} = 1$) where we will see that the dynamics is entirely determined by geometric considerations with the flow set by the site percolation threshold $p_c$ of the Lieb lattice, and (2) cases away from perfect swapping ($p_{hop}<1$) where both the stochastic nature of the dynamics and site percolation effects play a role.


\subsection{Perfect swapping Case: Percolation Threshold}

With $p_{hop} = 1$, the dynamics is deterministic, with particles moving in a chiral fashion along edges while performing localized trajectories in the bulk. Importantly, the chiral flow in this case is robust against geometric deformations as shown in our previous work \cite{wampler2021stirring}. Therefore the flow across the system given a specific disorder realization is simply determined by whether a percolating cluster can be formed, and the disorder average flow should exhibit a percolation threshold as function of the local vacancy probability $p_{\alpha}$. We numerically study the mean chiral flow per cycle as a function of the site vacancy probability $p_{\alpha}$ averaged over disorder realizations in the perfect swapping case. We denote chiral flow as $F$ and mean chiral flow averaged over disorder realizations as $\langle F \rangle$.

In Fig \ref{fig: perfect swapping percolation}(a), we set the linear system site to take values $L = 25, 33, 41$. We observe that at low disorder, the mean flow per cycle $\langle F \rangle$ is close to unity. Increasing the probability of site vacancy eventually causes the flow to drop sharply to zero chiral flow, with a sharper percolation threshold expected as we increase $L$. 

The disorder averaged $\langle F \rangle$ at different linear lattice sizes intersect around $p_c \approx 0.26$ in Fig \ref{fig: perfect swapping percolation}(a), roughly matching the peak fluctuations (standard deviation) in the chiral flow across disorder realization in Fig \ref{fig: perfect swapping percolation}(b). To understand the critical properties of the flow near percolation threshold in the perfect swapping case, we note that since F determines whether a percolating cluster exists in a given disorder configuration, having the values $1$ (if percolating cluster exists) or $0$ (if no percolating cluster exists). When taken as an average over different disorder configurations, $\langle F  (p_{\alpha}, p_{hop}=1) \rangle$ then gives the probability that a percolating cluster exists.

The percolation threshold $p_c$ we obtained matches with the earlier result on 2D site percolation in Lieb lattice in an earlier work \cite{PhysRevE.104.064122}, but there we should interpret the result by mapping $p_{\alpha} \rightarrow 1 - p$ as we starts with a filled Lieb lattice and adding in site vacancy when tuning up $p_{\alpha}$, while \cite{PhysRevE.104.064122} starts with empty Lieb lattice and gradually filling the lattice sites.

Given that disorder averaged $\langle F \rangle$ gives the probability that a concentration $p_{\alpha}$ gives a percolating cluster, according to the classic percolation theory \cite{percolationbook}, we then expect such quantity to exhibit finite size ansatz of the form $\langle F \rangle = f((p_{\alpha}-p_c) L^{1/\nu})$ to describe the percolation threshold around $p_c \approx 0.26$. We find that the critical exponent obtained is around $\nu \approx 1.35$, which roughly coincide with the 2D site percolating Lieb lattice result in \cite{PhysRevE.104.064122}. The fluctuations of the mean chiral flow across disorder realizations also peak around the percolation threshold $p_c \sim 0.26$, and the size of peak increasing with system length $L$ as shown in the middle panel in Fig \ref{fig: perfect swapping percolation}.

\subsection{Away from Perfect swapping: Crossover}

\begin{figure}[h]
    \centering
    \includegraphics[width=0.5\textwidth]{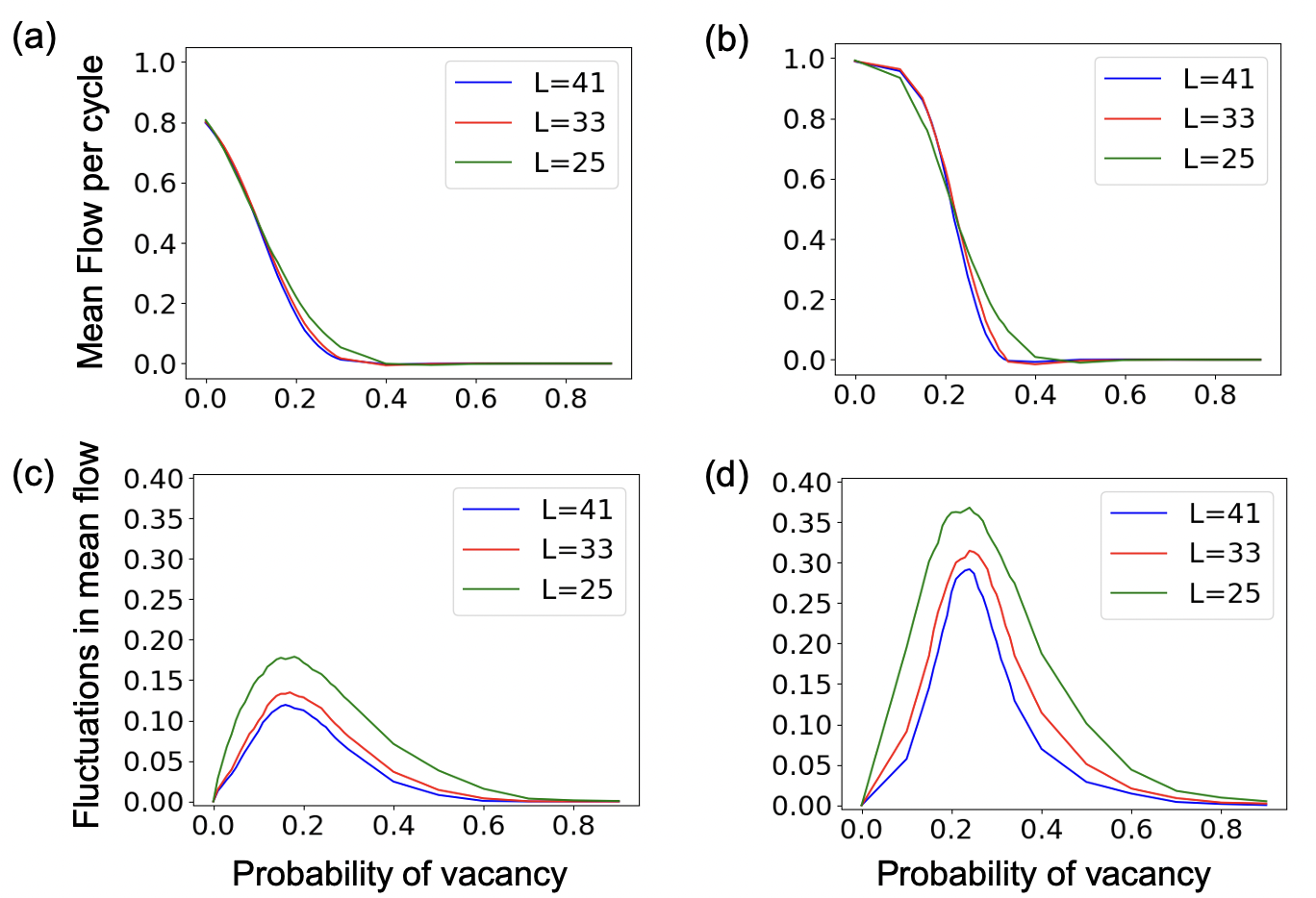}
    \caption{(a) Mean flow per cycle for $p_{hop} = 0.8$ and (b) for $p_{hop} = 0.95$, averaged over 1000 disorder configurations, (c) Fluctuations (standard deviation) of the mean flow per cycle for $p_{hop} = 0.8$ and (d) for $p_{hop} = 0.95$ over across 1000 disorder realizations for various system sizes $L$. Note that in contrast to the $p_{hop} = 1$ case, the fluctuation decreases as a function of system sizes, signifying a crossover. $p_{hop} < 0.8$ has similar qualitative features for the mean flow per cycle curve as that of $p_{hop} = 0.8$ and a less pronounced fluctuation.}
    \label{fig: finite hopping crossover}
\end{figure}

Away from the perfect swapping case, the flow ceases to be robust over a finite range of disorder and drops as soon as we introduce site dilution, as shown in Figure \ref{fig: finite hopping crossover} (a) and (b). The sharp transition feature we have observed in the perfect swapping case is therefore specific to the perfect swapping case, where we have deterministic walks rather than random walk for the case of finite hopping probability $p_{hop} < 1$, where now both random walks and site percolation affect the chiral flow rate in our study. 

Another distinction with the perfect swapping case can be seen from the fluctuations in the chiral flow rate across disorder realizations as shown in Figure \ref{fig: finite hopping crossover} (c) and (d). In contrast with the perfect swapping case (where fluctuation of the flow increases with system size), the fluctuation of the flow decreases as we increase the system size. The averaged flow across disorder no longer exhibits sharp percolating threshold, unlike the perfect swapping case. 

\begin{figure}[h]
    \centering
    \includegraphics[width=0.5\textwidth]{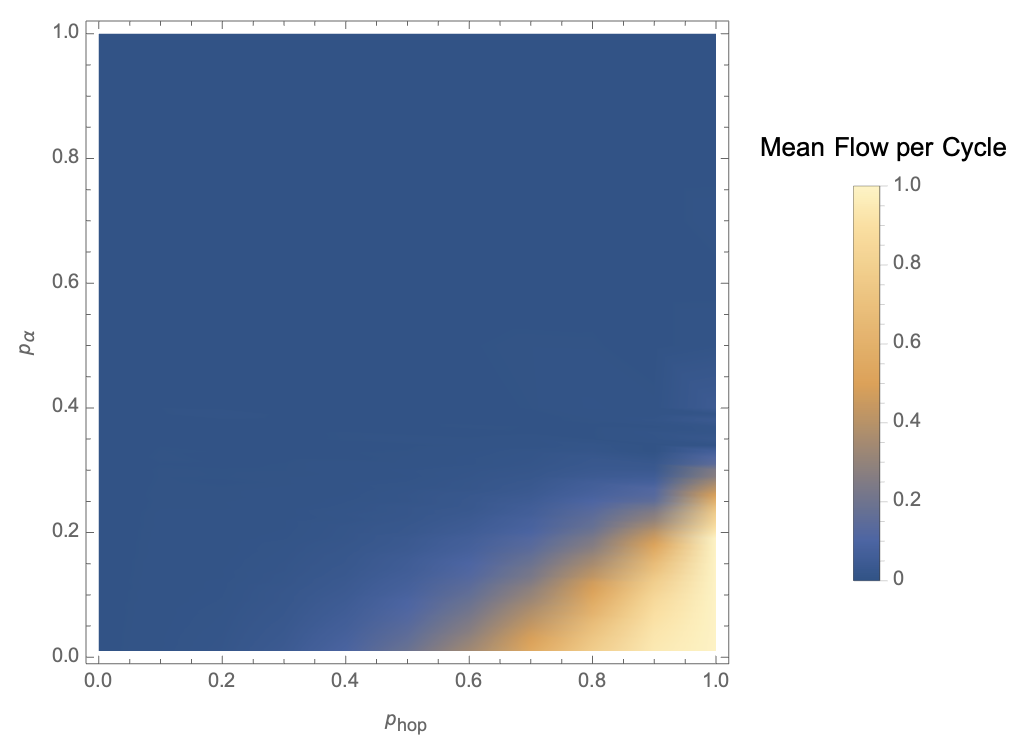}
    \caption{2D plot of the mean chiral flow per cycle for site vacancy for a $33 \times 33$ Lieb lattice, averaged over 1000 disorder realizations. The horizontal axis represents hopping probability $p_{hop}$ and the vertical axis is the probability of vacancy in the lattice $p_{\alpha}$. The percolation transition happens on the line $p_{hop} = 1$.}
    \label{fig: 2d plot phop vs palpha}
\end{figure}

The physical reason for the difference is that the flow is a quantity that is affected by both the hopping probability and site vacancy, and in the case of random walk $p_{hop} < 1$ both factors affect the late time chiral flow per cycle, obscuring contributions from site vacancy (geometric percolation) alone. This is not an issue for perfect swapping case, where the only factor affecting the flow is geometrical. Finally, we present a contour plot of the mean chiral flow per cycle across disorder realizations with 2 parameters of interest here in Figure \ref{fig: 2d plot phop vs palpha}: the hopping probability $p_{hop}$ and the probability of blockade disorder $p_{\alpha}$. 

A comment is in order for stochastic random walks on lattice with site vacancies. In \cite{gefenanomalousdiffusion}, particle diffusion was studied in the context of site percolation. When no percolating cluster can be formed (in our convention, $p_{\alpha} > p_c$), the mean square spreading $\langle r^2 \rangle = \text{const}$ in late time dynamics for a particle undergoing diffusive behaviour. In another limit when percolating cluster can always be formed ($p_{\alpha} < p_c$), we have normal diffusive behaviour $\langle r^2 \rangle = D t$ in late time. However, interesting anomalous diffusive behaviour occurs right at $p_{\alpha} = p_c$ according to \cite{gefenanomalousdiffusion, percolationbook}, where $\langle r^2 \rangle \propto t^{2/3}$. While we have not investigated diffusive transport for the case of site vacancy disorder, this will constitute an interesting point to investigate.

\section{Lattice Distortion and Onsite Potential disorder}
\label{Sec: lattice distortion and on-site disorder}

We next investigate two additional models of disorder.  Namely, we consider the case where the hopping parameter strength between sites is disordered and the case where a random on-site potential is applied to each site (represented in Figs. \ref{fig: disorder summary}c and \ref{fig: disorder summary}b respectively with explicit details for each model below).  We will again restrict ourselves to the Zeno limit, leaving results on the effects of disorder away from the Zeno limit to Section \ref{Sec: Disorder Away From Zeno}.      


The first model we consider is the application of the measurement protocol to a Lieb lattice with the random hopping Hamiltonian 
\begin{equation}
    H = - \sum_{\langle \mathbf{r} \mathbf{r'}  \rangle} t_{ \mathbf{r} \mathbf{r'}} a_{\mathbf{r}}^{\dagger} a_{\mathbf{r'}}, 
\end{equation}
where $t_{ \mathbf{r} \mathbf{r'}}$ is now a random variable drawn from the uniform random distribution $[ -\delta t + 1, \delta t + 1  ]$, i.e. we set the mean of $\langle t_{ \mathbf{r} \mathbf{r'}} \rangle = 1$ with disorder strength $\delta t$, which we allow to be at max $|\delta t| \leq 1$.  Hopping disorder is associated with random lattice distortion, with distances between lattice sites randomly lengthened or shortened, leading to an alteration in the hopping integral $t_{\text{hop}}$ between sites. 

In the Zeno limit, we calculate the transition matrix $R_i$ which characterizes the evolution for the $i^{th}$ step of the measurement protocol with the random hopping disorder, replacing  Eq. \eqref{eq: stochastic} with 
\begin{gather}
    R_i=\oplus_{\langle \mathbf{r} , \mathbf{r'} \rangle\in  A_i}\left(
    \begin{array}{cc}\small
        1-p_{hop,\mathbf{r} \mathbf{r'}}  &  p_{hop,\mathbf{r} \mathbf{r'}}  \\
        p_{hop,\mathbf{r} \mathbf{r'}}  & 1-p_{hop,\mathbf{r} \mathbf{r'}} \\
    \end{array}
    \right) \oplus_{\text{other sites}} I  
    \label{eq: R form}
\end{gather}
and the hopping probability depends on $t_{\mathbf{r} \mathbf{r'}}$ via
\begin{eqnarray*}
    p_{hop, \mathbf{r} \mathbf{r'}} = \sin^2(\frac{t_{\mathbf{r} \mathbf{r'}} T}{8}) . 
\end{eqnarray*} 
In other words, the random hopping Hamiltonian translates to a periodic random walk where different links have different hopping probabilities. 

\begin{figure}[h]
    \centering
    \includegraphics[width=0.45\textwidth]{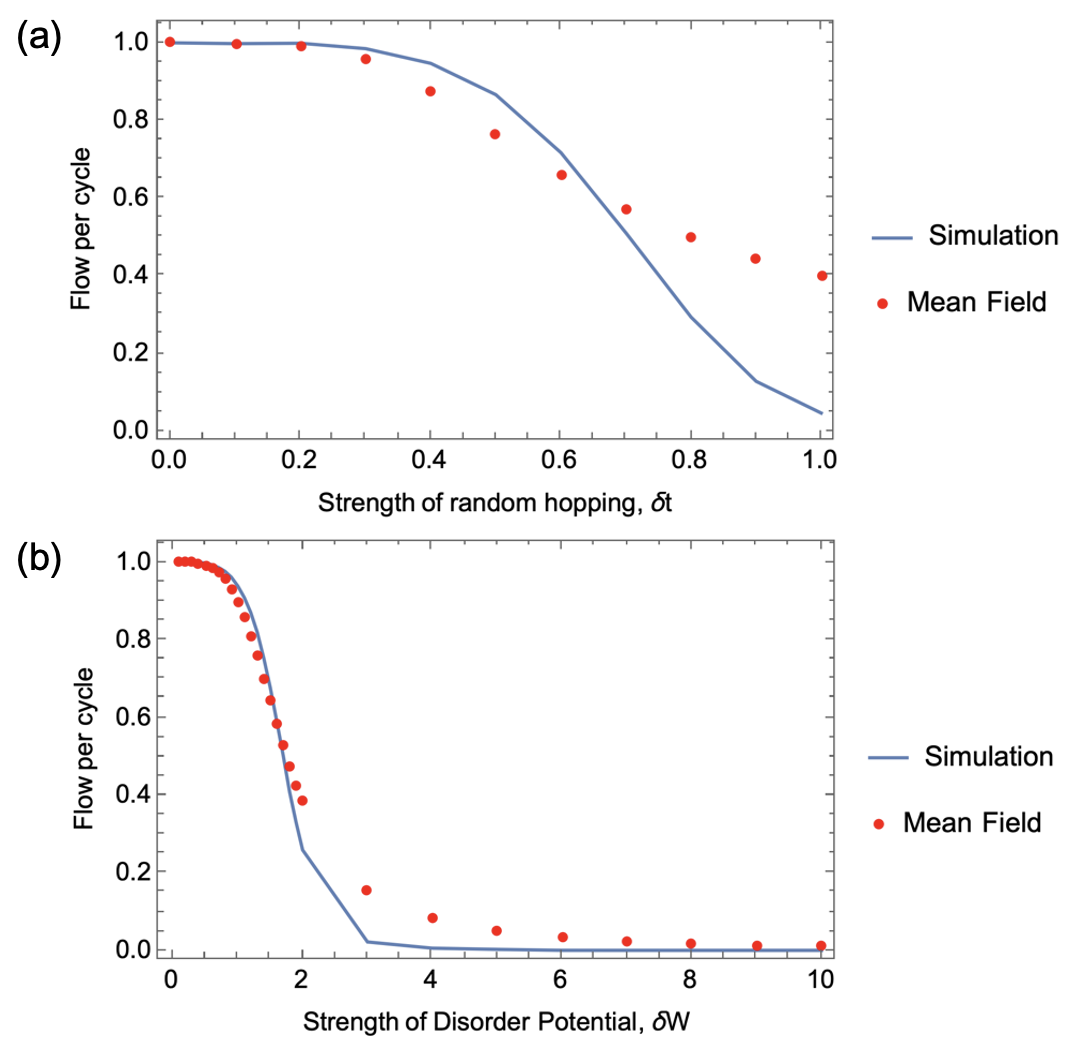}
    \caption{(a) The mean Flow per cycle for the case of random hopping strength $\delta t$ and (b) The mean Flow per cycle for the case of random potential $\delta W$. Here the line represents numerical simulation taken on a $33 \times 33$ Lieb lattice averaged over 1000 disorder realization, and is compared to the mean field result (dot) outlined in the main text.}
    \label{fig: Disorder Zeno Mean Field vs Numerics}
\end{figure}

\begin{figure}[h]
    \centering
    \includegraphics[width=0.45\textwidth]{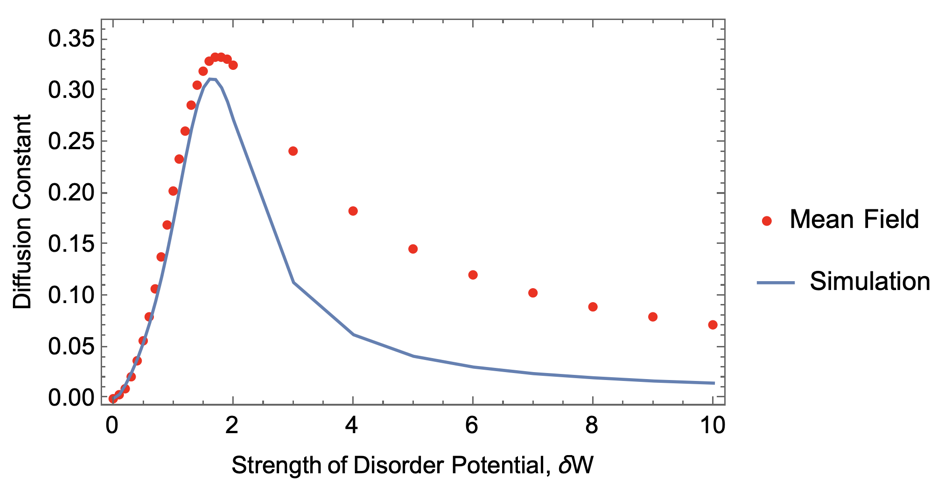}
    \caption{The mean diffusion constant as a function of the strength of disorder potential for system with random onsite potential averaged over 1000 disorder realizations in the Zeno limit. Here $T = 4 \pi$ corresponds the perfect swapping case in the clean limit without any diffusion. The mean diffusion constant from the crude mean field treatment agrees with the numerical simulations for weak disorder.}
    \label{fig: diffusion zeno potential}
\end{figure}

The second disorder model we study in this section is a random on-site potential. We note that it was shown in \cite{Titum_2015} that this variety of disorder, when added to the Floquet model of Rudner et. al. \cite{rudner2013anomalous}, may prevent bulk diffusion (due to Anderson localization) while still preserving protected edge transport.  Such a system is referred to as an anomalous Floquet-Anderson Insulator.  

We note that, in the measurement-induced model, the diffusive behavior in the bulk is expected even when a disordered potential is present. This occurs even if the disorder is sufficient to result in Anderson localization in the absence of measurements. Indeed, Anderson localization is a wave effect that emerges due to interference within a single particle wave function as it moves through a random potential. The projective measurements rapidly collapse the wave function to single sites within the lattice, thus no interference is possible ruining Anderson localization (as long as the distance between measured sites is smaller than the expected localization length). We note, however, that Anderson localization physics may still play a role in the limit where the measurements are temporally and spatially sparse enough \cite{popperl2022measurementAnderson}.     

In this section, we analyze in detail the effect of the disorder on the diffusion constant as well as its effect on the chiral edge flow. We also present a mean-field type argument which approximately captures these effects. 
 

The Hamiltonian for the case with random on-site potential takes the form 
\begin{equation}
    H = - t_{hop} \sum_{\langle \mathbf{r} \mathbf{r'} \rangle} a_{\mathbf{r}}^{\dagger} a_{\mathbf{r'}} + \sum_{\mathbf{r}} W_{\mathbf{r}} a_{\mathbf{r}}^{\dagger} a_{\mathbf{r}}, 
\end{equation}
where $W_{\mathbf{r}}$ is sampled from the random uniform distribution $W_{\mathbf{r}} \in [ - \delta W, \delta W ]$. We set $t_{hop} = 1$ and $T = 4 \pi$ in our numerical results below, while we keep the variable $t_{hop}$ and $T$ in the our expression of the random stochastic transition matrix below (Eqn \ref{eq: disorder onsite stochastic}). The random stochastic transfer matrix describing two neighboring unmeasured sites $\mathbf{r}$ and $\mathbf{r'}$ is (surrounded by measured sites and in the Zeno limit) takes the form with $R_i$ from Eqn \ref{eq: R form} with $p_{hop, \mathbf{r} \mathbf{r'}}$ taking the form below instead (for derivation see Appendix \ref{Appendix: Onsite}) 
\begin{eqnarray}  
    p_{hop, \mathbf{r} \mathbf{r'}} = \frac{2 t_{hop}^2 (1 - \cos( \frac{T}{8}\sqrt{4 t_{hop}^2 + (W_{\mathbf{r}} - W_{\mathbf{r'}} )^2 } ) )}{4 t_{hop}^2 + (W_{\mathbf{r}} - W_{\mathbf{r'}})^2} . \label{eq: disorder onsite stochastic}
\end{eqnarray}
Let us unpack Eq \ref{eq: disorder onsite stochastic}. For low disorder, the typical chemical potential difference between adjacent lattice sites is small and the model is close to the mean hopping probability $p_{hop}$ determined by $T$. As disorder strength increases, the denominator grows while the numerator stays bounded by the cosine function, leading to the hopping probability to zero (thereby freezing the dynamics) due to huge potential difference.

Let us interpret the numerical results on the mean chiral flow per cycle and diffusion constant for the case of random onsite potential as shown in Fig \ref{fig: Disorder Zeno Mean Field vs Numerics} (b) and \ref{fig: diffusion zeno potential} respectively. In the intermediate disorder strength, we have an interesting situation where chirality is partially suppressed but diffusion transport proliferates as shown in Fig \ref{fig: Disorder Zeno Mean Field vs Numerics} and \ref{fig: diffusion zeno potential} respectively. This can be understood in that in the intermediate disorder strength we have a random bond model where many links takes intermediate hopping probability $0 < p_{hop, \mathbf{r} \mathbf{r'}} < 1$. In the intermediate disorder case with random walk, diffusion is more pronounced than in the case close to perfect swapping (with deterministic walk) or freezing (with $p_{hop} \approx 0$). In the strong disorder case, we again have strong localization in both the measurement induced chiral transport and diffusive transport.

\begin{figure*}[t]
    \centering
\includegraphics[width=1\textwidth]{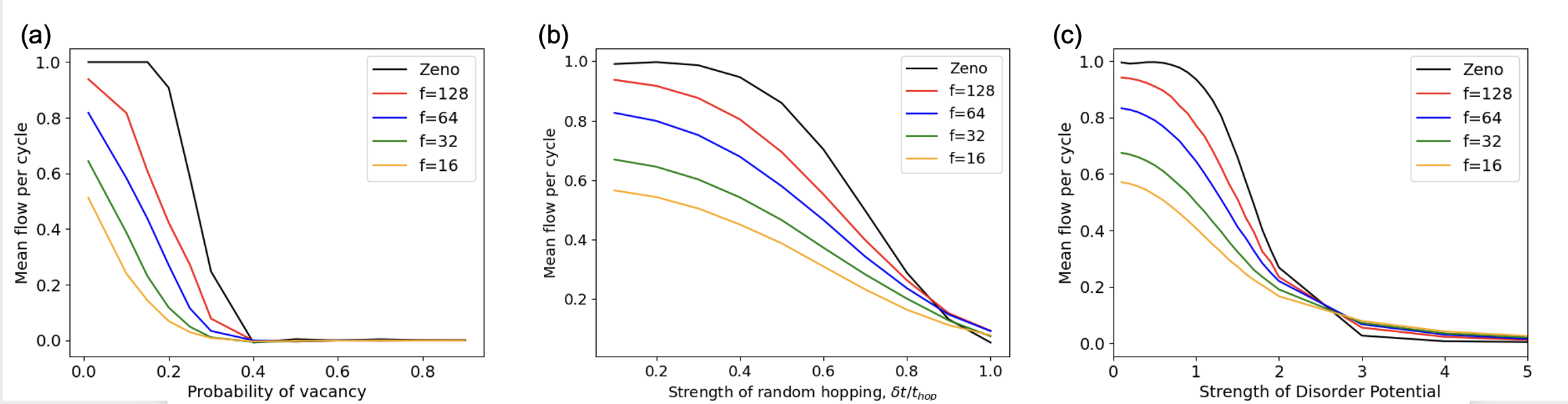}
    \caption{Comparison between flow per cycle for the case of (a) site vacancy, (b) random hopping strength, and (c) disordered potential. The comparisons are made across different measurement frequencies against the Zeno limit case and in all cases the lattice size is $33 \times 33$ and averaged across 1000 disorder realizations.}
    \label{fig: disorder away zeno}
\end{figure*}

In the limit of weak disorder, we can estimate the flow by computing the spatial average of $p_{hop, \mathbf{r}\mathbf{r'}}$ at different links and using the resulting spatially averaged $\bar{p}_{hop}$ as the effective hopping probability in Eqn (\ref{eqn: F_analytic}) to compute the flow transport in a translationally invariant system to get the effective flow. In this sense, we call this approach the 'mean-field' approach, where we replace a disordered model with inhomogeneous hopping probability $p_{hop, \mathbf{r}\mathbf{r'}}$ at different links with homogeneous $\bar{p}_{hop}$ over all links, and treat it as if it has translational invariance. The inhomogeneity in the hopping probability at different links in a disordered model can be caused by random hopping strength or random onsite potential.

For the case of random hopping, the effective hopping $\bar{p}_{hop}$ is given by averaging
\begin{eqnarray}
    \bar{p}_{hop} (\delta t, t_{hop}) = \int_{t_{hop} - \delta t}^{t_{hop} + \delta t} dt_{\mathbf{r} \mathbf{r'}} \sin\left( \frac{t_{\mathbf{r} \mathbf{r'}} T}{8} \right).
\end{eqnarray}
We compute $F$ using the translational invariant formula Eqn (\ref{eqn: F_analytic}) using  $\bar{p}_{hop}$ for various $\delta t$ and compare with direct numerical simulations in Fig \ref{fig: Disorder Zeno Mean Field vs Numerics} (a). As expected, the agreement goes well for small and intermediate $\delta t \sim 0.3$ relative to $t_{hop} = 1$ before deviation occurs for larger $\delta t$. 

In the presence of potential disorder, the average hopping probability as function of disorder strength $\delta W$ and hopping $t_{hop}$ is 
\begin{eqnarray}
    \bar{p}_{hop} (\delta W, t_{hop}) = \frac{1}{4 \delta W^2} \int_{-\delta W}^{\delta W} \int_{-\delta W}^{\delta W} dW_{\mathbf{r}} dW_{\mathbf{r'}} \times \nonumber \\  \frac{2 t_{hop}^2 (1 - \cos( \frac{T}{8}\sqrt{4 t_{hop}^2 + (W_{\mathbf{r}} - W_{\mathbf{r'}} )^2 } ) )}{4 t_{hop}^2 + (W_{\mathbf{r}} - W_{\mathbf{r'}})^2}.
\end{eqnarray}
In Fig. \ref{fig: Disorder Zeno Mean Field vs Numerics}(b) we show the effective 'mean-field' result vs numerical simulations of the disordered system, and similarly find good agreement at weak disorder. Note that both numerically and mean field approximation show that large disorder leads to an effective suppression of flow.


We also performed the 'mean-field' approach to evaluate the diffusion constant for the case of random onsite potential in Fig \ref{fig: diffusion zeno potential}. In this case, the diffusion constant for the mean-field approach agrees with numerical simulation for weak disorder, but is not as tight for intermediate disorder compared to its use in for estimating the flow \ref{fig: Disorder Zeno Mean Field vs Numerics}. It is interesting to note that the maximal diffusion coefficient coincides with the region where the drop in chiral flow as function of disorder strength is the steepest.

We emphasize that the actual flow depends in a complicated nonlinear way on the particular disorder realization, therefore we can only expect the above approach to work in the weak disorder limit, where fluctuations in $ p_{hop, \mathbf{r} \mathbf{r'}}$ are small. 

In the next section we turn to consider all variety of disorders considered in this paper away from the Zeno limit in the next section.

\section{Numerical Results Away from Zeno Limit}
\label{Sec: Disorder Away From Zeno}

We now investigate the effect of relaxing the Zeno limit assumption on our measurement protocol in the chiral flow transport in various disordered systems we previously simulated. In Figure \ref{fig: disorder away zeno}, we study the cases of site vacancy, random hopping strength and random onsite potential against finite measurement frequencies of 16, 32, 64, and 128 measurements per measurement step (8 measurement steps make up a measurement cycle in our protocol). 

Intuitively, the overall effect of reducing measurement frequency will tend to reduce the amount of chiral flow transport in our systems, as shown in Fig \ref{fig: disorder away zeno}. In Figure \ref{fig: disorder away zeno} (a), we study site vacancy disorder where we set $T = 4 \pi$, which in the Zeno limit corresponds to perfect swapping. The robustness of the chiral flow to low disorder in the Zeno limit disappears as soon as we tune the measurement frequency away from the Zeno limit and the chiral flow starts decreasing as soon as we introduce disorder. The same feature of the lack of robustness against (global) minute disorder are also seen with the random hopping strength in Fig \ref{fig: disorder away zeno} (b) and random onsite potential in Fig \ref{fig: disorder away zeno} (c). 

Nonetheless, we would like to note the interesting case of random onsite potential in Fig \ref{fig: disorder away zeno}(c). Near the clean limit, we expect that lowering the measurement frequency will lower the chiral flow. Meanwhile, strong disorder limit suppresses both diffusive and chiral transport in our half-filled system. In a finite window of intermediate onsite disorder strength, however, decreasing the measurement frequency actually enhances the chiral flow. However, whether this effect is an artifact of finite size/time effect, or a genuine non-trivial effect arising from the interplay between chiral flow and diffusive spreading, is currently unknown and this is a possible avenue for future work.

\begin{figure}[h]
    \centering
    \includegraphics[width=0.45\textwidth]{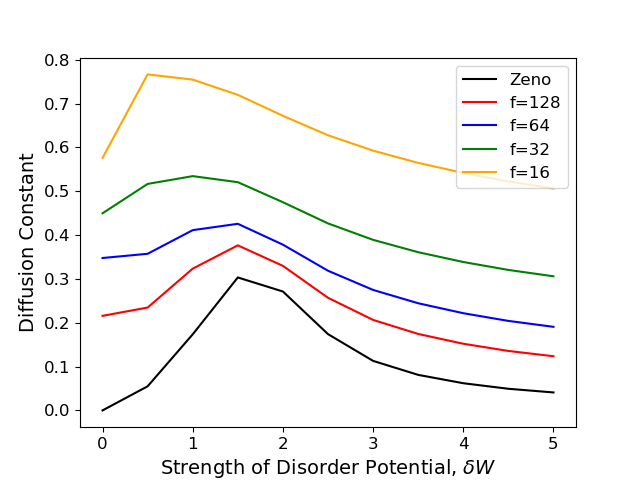}
    \caption{Diffusion constant for various measurement frequencies (and the previous Zeno limit result in \ref{fig: diffusion zeno potential}) as a function of disorder potential strength $\delta W$, simulated with $T = 4 \pi$ over 1000 disorder realizations.}
    \label{fig: diffusion potential away zeno}
\end{figure}

Finally, we turn our attention to diffusive dynamics, specializing in the case of diffusion constant of the measurement-induced chiral system under random onsite potential away from the Zeno limit. The diffusion constant generally increases as the measurement frequency is decreased away from the Zeno limit, which aligns with the intuition that away from the Zeno limit particle dynamics in our measurement protocol becomes more diffusive. The result is presented in Fig \ref{fig: diffusion potential away zeno}, where we fix $T = 4 \pi$ for all measurement frequencies.

\section{Discussions and Outlook}

In this work we presented a systematic investigation on the diffusive dynamics and the effect of disorder on measurement-induced chirality exhibited by free fermions under various disorder types. 
In particular, there is a putative percolation transition-like behavior exhibited by the mean chiral flow of the particles in site blockade disorder in the disorder limit. It is also noteworthy that our measurement protocol in general is reliably robust to the introduction to global disorder, with mean chiral flow rate decreasing significantly only when disorder strengths are significant. Finally, we also provided various analytical mean field picture to describe the random hopping and random onsite potential cases and the agreement holds up to significant disorder.

It is an interesting question whether the behavior we find persists if the disorder were time dependent. In standard quantum systems, a time-dependent disorder is fundamentally different from a quenched disorder (e.g. Anderson localization is absent in the dynamic case \cite{Segev-Fishman}). Because of the time-dependent protocol used for measurement-induced chirality, we speculate that dynamic disorder should result, in the same behavior as the quenched disorder we consider above, at least in the Zeno limit. However, as the rate of measurement drops, corrections due to coherent behavior and localization phenomena may start appearing at intermediate times. The investigation of time dependent random potentials in the low measurement frequency limit would be an interesting direction for additional work.

Several comments are in order regarding the relation of our work to recent relevant literature. This work analyzed the average transport and dynamics of densities over all possible measurement outcomes and averaged over different disorder realizations. The recent work by P\"{o}pperl \textit{et. al.} \cite{popperl2022measurementAnderson} studies particular quantum trajectories (measurement outcomes) of the particle density profile using wavefunctional approach (while we used density matrices instead) with different interesting averages for a 1D fermion on an Anderson localized chain. \cite{szyniszewski2022disorderedmonitor} also studied free fermions on 1D Anderson chain, but has focused primarily on entanglement properties.

We now turn to discuss open problems and possible avenue for future work in the general direction of measurement-induced chirality. First, it would be interesting to define an effective "cyclotron frequency" $\omega_{eff} \sim ~1/T$ which is proportional to an effective magnetic field $B_{eff}$  for this kind of system. This is inspired by analogy of our measurement-induced chiral flow to the anomalous Floquet topological insulator proposed in \cite{rudner2013anomalous} and would constitute an interesting subproblem to develop the idea further.

Various interesting investigations can also be further explored on the diffusive dynamics in for monitored fermions exhibiting measurement-induced chirality. It will be interesting to investigate to see if statements can be made about how the diffusion coefficient in the diffusion dynamics from an occupied region to the empty in our setup can be related to Fick's law. Another interesting thought to exploration is the question of describing an effective electrical resistance for our system given the diffusion constants for our system.

Finally, one could also extend the question of measurement-induced transports to measurement-induced delocalization transitions \cite{popperl2022measurementAnderson} and other interesting quantum walk behaviours \cite{measurementwalk2022, measurementwalkIBM2022} caused by different types of measurements. The direction of using measurement to engineer interesting transport and dynamics is a nascent and new area of research that could potentially lead to more interesting discoveries.

\section*{Acknowledgments}

The work of I.K., B.J.J.K. and M.W. was supported in part by the NSF grant DMR-1918207. G.R. acknowledges support  from  the  Institute  of  Quantum Information  and  Matter,  an  NSF  Physics  Frontiers  Center funded  by  the  Gordon  and  Betty  Moore  Foundation, and the Simons Foundation.

\bibliography{main.bib}
\bibliographystyle{unsrt}

\onecolumngrid
\appendix

\section{Derivation for the analytical expression for the Diffusion constant} \label{Appendix: Diffusion}

\begin{figure}[h]
    \centering
    \includegraphics[width=0.4\textwidth]{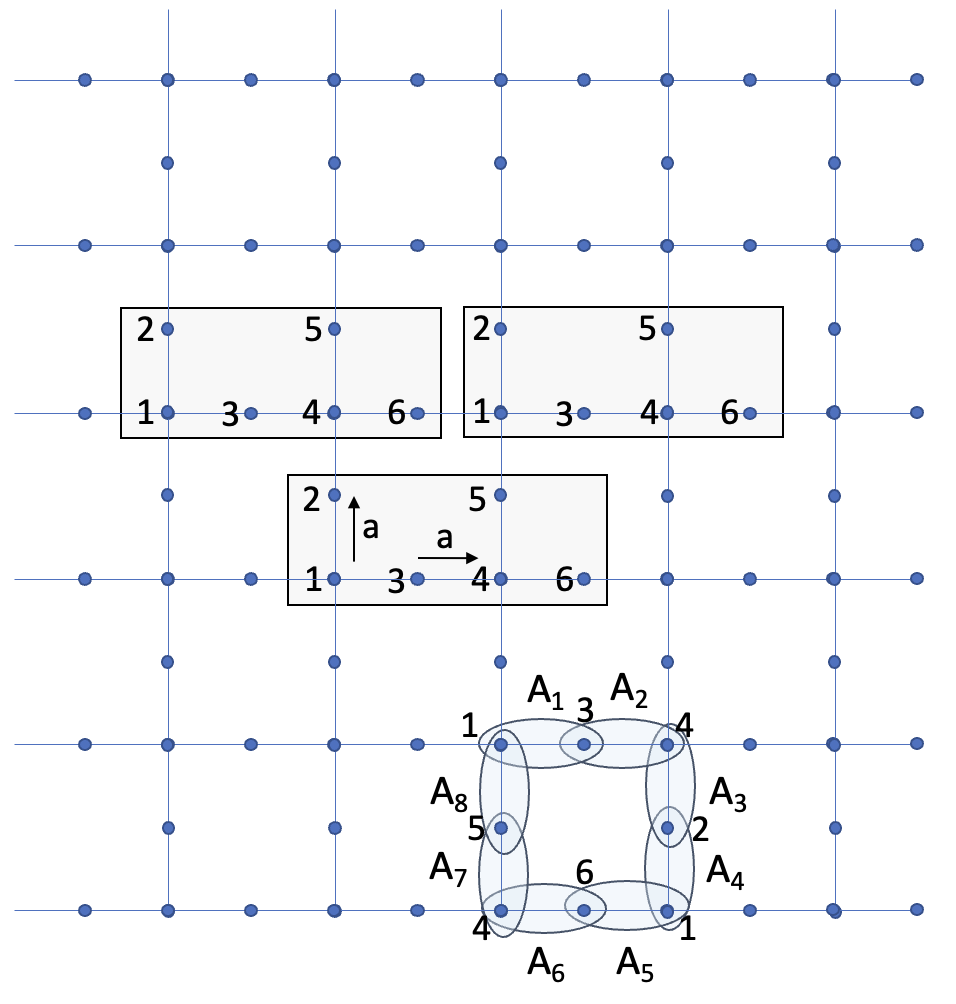}
    \caption{The Lieb lattice is divided into 6 lattice sites per unit cell, and the set of unmeasured sites in each step is shown as $A_i$ in the Figure. Here, we note that we count the distance between neighbouring lattice sites with lattice constant $a$ (conveniently set to 1) rather than counting that as the inter-unit cell distance in typical systems for consistency with numerical simulation when computing the diffusion constant $D$.} 
    \label{fig: Setup Analytic Diffusion}
\end{figure}

In this Appendix, we derive the expression Eq \ref{eq: diffusion k space} from the definition Eq \ref{eq: diffusion def} using a translationally invariant setup in Figure \ref{fig: Setup Analytic Diffusion} and working in momentum space. The setup is shown in Figure \ref{fig: Setup Analytic Diffusion}.

We define a set of consistent Fourier transformation by using the following conventions
\begin{eqnarray}
    R_{cyc} (\mathbf{r}, \mu; \mathbf{r'}, \nu) = \int \frac{d^2 k}{(2 \pi )^2} R_B (\mathbf{k}, \mu, \nu) e^{i \mathbf{k} (\mathbf{r} - \mathbf{r'})}  \label{eq: R Fourier r to k} \text{ , } R_B (\mathbf{k}, \mu, \nu) = \sum_{(\mathbf{r}-\mathbf{r'})} e^{- i \mathbf{k} (\mathbf{r} - \mathbf{r'})} R_{cyc} (\mathbf{r}, \mu; \mathbf{r'}, \nu) 
\end{eqnarray}
where $V = L_x L_y$. Here, we would like to make two remarks about our convention. (1) The matrix $R_{cyc} (\mathbf{r}, \mu; \mathbf{r'}, \nu)$ is only dependent on the difference $\mathbf{r} - \mathbf{r'}$, by making use of the translational invariance, hence Eq \ref{eq: R Fourier r to k}. (2) We count distance a little differently than how one would normally count distance for setup with unit cell decomposition. Normally, one only keeps track of the distance between different unit cell in the expression $(\mathbf{r} - \mathbf{r'})$ in Eq \ref{eq: R Fourier r to k}. However, as our numerical calculation keeps track of distance between lattice sites rather than unit cells, we will do likewise for our $k$-space calculation for consistency.

For example, the $k$-space $R_5 (\mathbf{k},\theta)$ and $R_4 (\mathbf{k}, \theta) $ are $6 \times 6$ matrices of the following form respectively
\begin{eqnarray*}\small
    R_{5}(\mathbf{k}, \theta)=\left(
    \begin{array}{cccccc}
    1-p & 0 & 0  & 0 & 0 & p e^{- i k_x} \\
    0 & 1 & 0 & 0 & 0 & 0 \\
    0 & 0 & 1 & 0 & 0 & 0 \\
    0 & 0 & 0 & 1 & 0 & 0 \\
    0 & 0 & 0 & 0 & 1 & 0 \\
    p e^{i k_x} & 0 & 0 & 0 & 0 & 1-p 
    \end{array}
    \right) \text{ , } 
    R_{4}(\mathbf{k}, \theta)=\left(
    \begin{array}{cccccc}
    1-p & p e^{i k_y} e^{i \theta} & 0  & 0 & 0 & 0 \\
    p e^{-i k_y} e^{-i \theta} & 1-p & 0 & 0 & 0 & 0 \\
    0 & 0 & 1 & 0 & 0 & 0 \\
    0 & 0 & 0 & 1 & 0 & 0 \\
    0 & 0 & 0 & 0 & 1 & 0 \\
    0 & 0 & 0 & 0 & 0 & 1 
    \end{array}
    \right)
\end{eqnarray*} 
and we can construct $R_B (\mathbf{k}, \theta) = R_8 R_7 R_6 R_5 R_4 R_3 R_2 R_1 $ based on this construct of the $k$-space stochastic transition matrices by keeping track of the factors $e^{-i k_x}$ whenever the particle hops to the right and $e^{- i k_y} e^{-i \theta}$ whenever the particle hops upward, and vice versa for the left and downward hopping elements, and $\theta$ is the counting field keeping track of the vertical flow alone in our setup.

Starting from Eq \ref{eq: diffusion def}, which we reproduce here for convenience,
\begin{eqnarray}
    &\mathbf{r}_{mean} (t=8N) \equiv \frac{\sum_{\mathbf{r}} \mathbf{r} G_{\mathbf{r} \mathbf{r}} (t=8N) }{\sum_{\mathbf{r}} G_{\mathbf{r} \mathbf{r}} (t=8N)} , \nonumber \\
    &\langle \Delta r^2 (t=8N) \rangle \equiv \frac{\sum_{\mathbf{r}} ( \mathbf{r} - \mathbf{r}_{mean} (t=8N) )^2 G_{\mathbf{r} \mathbf{r}} (t=8N) }{\sum_{\mathbf{r}} G_{\mathbf{r} \mathbf{r}} (t=8N)} , \nonumber \\
    &D = \lim_{N \rightarrow \infty} \frac{\langle \Delta r^2 (t=8N) \rangle - \langle \Delta r^2 (t=0) \rangle }{8N}
\end{eqnarray}
For the term $\sum_{\mathbf{r}} ( \mathbf{r} - \mathbf{r}_{mean} )^2 G_{\mathbf{r} 
\mathbf{r}} (t=8N)$ term, we simplify to get 
\begin{eqnarray}
     \sum_{\mathbf{r}} ( \mathbf{r} - \mathbf{r}_{mean} )^2 G_{\mathbf{r} \mathbf{r}} (t=8N) = \sum_{\mathbf{r}}  r^2  G_{\mathbf{r} \mathbf{r}} (t=8N) - r^2_{mean} (t=8N)
\end{eqnarray}

We Fourier transform the real space $G_{\mathbf{r'} \mathbf{r'}} (t=8N) \equiv ( R_{cyc}^N )_{\mathbf{r'} \mathbf{r}} \ket{g_{\mathbf{r}} (t=0)}$ according to Eq \ref{eq: R Fourier r to k} in the definition $\langle \Delta r^2 \rangle$. In our current setup, we start with a single particle with unit density placed on the origin so that $\langle \Delta r^2 (t=0) \rangle=0$, and since there is no injection and extraction, particle number is conserved and we have $\sum_{\mathbf{r}} G_{\mathbf{r} \mathbf{r}} = 1$ at all times. Focusing on the $\sum_{\mathbf{r}} r^2 (R_{cyc}^N)_{\mathbf{r} \mathbf{r'}} \ket{g_{\mathbf{r'}} (t=0)}$ term, we have
\begin{eqnarray}
     \sum_{\mathbf{r}, \mu} r^2 (R_{cyc}^N)_{(\mathbf{r}, \mu) ,  (\mathbf{r'}, \nu)} \ket{g_{\mathbf{r'}, \nu} (t=0)} &=& \sum_{\mathbf{r}, \mu} r^2 (R_{cyc}^N)_{(\mathbf{r}, \mu) ,  (0, 1)} = \sum_{\mathbf{r}, \mu} \int \frac{d^2 k}{(2 \pi)^2} r^2 e^{i \mathbf{k} \mathbf{r}} ( R_{B}^N (\mathbf{k}) )_{\mu, 1} \nonumber \\ 
     &=& \sum_{\mathbf{r}, \mu} \int \frac{d^2 k}{(2 \pi)^2} (- \nabla^2_{\mathbf{k}} e^{i \mathbf{k} \mathbf{r}}) ( R_B^N (\mathbf{k}) )_{\mu, 1} = \sum_{\mathbf{r}, \mu} \int \frac{d^2 k}{(2 \pi)^2} e^{i \mathbf{k} \mathbf{r}} ( - \nabla^2_{\mathbf{k}} R_B^N ( \mathbf{k}) )_{\mu, 1} \nonumber \\ 
     &=& \sum_{\mu} \int d^2 k \delta^2 (\mathbf{k}) (- \nabla_{\mathbf{k}}^2 R_B^N (\mathbf{k}) )_{\mu, 1} = \sum_{\mu=1}^6 [- \nabla_{\mathbf{k}}^2 R_B^N (\mathbf{k}) |_{\mathbf{k} = 0} ]_{\mu, 1} . \label{eq: first diffusion term}
\end{eqnarray}
In the first line, we make use of the fact that $\ket{g_{\mathbf{r'} , \nu} (t=0)} = \delta_{\mathbf{r'}, 0} \delta_{\nu, 1}$. In the second line, we use the fact that $r^2 e^{i \mathbf{k} \mathbf{r}} = - \nabla_{\mathbf{k}}^2 e^{i \mathbf{k} \mathbf{r}}$ and we integrate by part. From the second to the third line, we summed over $\mathbf{r}$ with $\sum_{\mathbf{r}} e^{- i \mathbf{k} \mathbf{r}} = (2 \pi)^2 \delta^2 (\mathbf{k})$ and finally we integrate with the delta function to arrive at our final expression.

For the $\mathbf{r}_{mean}^2$ term, we perform similar computation as the one above to get
\begin{eqnarray}
     r_{mean}^2 &=& \left[ \sum_{\mathbf{r}, \mu} \mathbf{r} (R_{cyc}^N)_{(\mathbf{r}, \mu), (0, 1)} \right] \cdot \left[ \sum_{\mathbf{r'}, \mu'} \mathbf{r} (R_{cyc}^N)_{(\mathbf{r'}, \mu'), (0, 1)} \right] \nonumber \\ 
     &=& \left[ \sum_{\mu=1}^6 [- i \nabla_{\mathbf{k}} R_B^N (\mathbf{k}) |_{\mathbf{k}=0} ]_{\mu,1} \right] \cdot \left[ \sum_{\mu'=1}^6 [- i \nabla_{\mathbf{k'}} R_B^N (\mathbf{k'}) |_{\mathbf{k'}=0} ]_{\mu',1} \right] \nonumber \\ 
     &=& - \left[ \sum_{\mu=1}^6 [ \partial_{k_x} R_B^N (\mathbf{k}) |_{\mathbf{k}=0} ]_{\mu,1} \right]^2 - \left[ \sum_{\mu=1}^6 [ \partial_{k_y} R_B^N (\mathbf{k}) |_{\mathbf{k}=0} ]_{\mu,1} \right]^2 . \label{eq: second diffusion term}
\end{eqnarray} 
Collecting both terms Eq \ref{eq: first diffusion term} and Eq \ref{eq: second diffusion term}, we get Eq \ref{eq: diffusion k space}.

\section{Extraction of the late time dynamics of the mean flow per cycle on a finite size lattice} \label{Appendix: Finite Size Late Time}

\begin{figure}[h]
    \centering
    \includegraphics[width=0.4\textwidth]{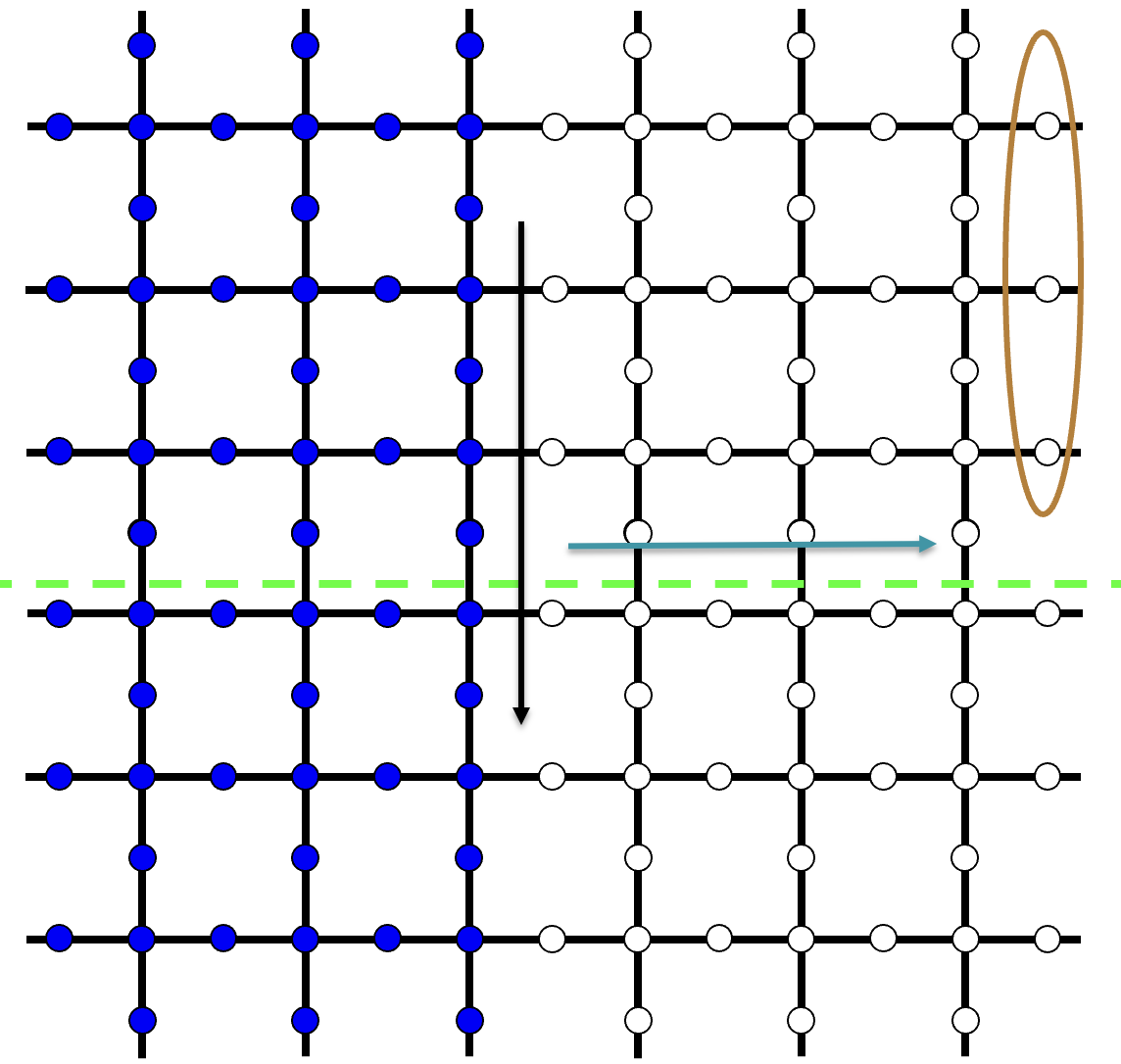}
    \caption{In our measurement protocol, the net chiral flow is measured by the number of particles transported along the direction of the black arrow. The diffusive transport takes place along the transverse direction (the direction of the grey arrow). We constantly inject particle on the left edge of the lattice and extract particle on the right edge. We keep track of the particle density at the top half edge of the lattice (circled sites) prior to extraction to truncate and obtain the late time chiral flow per cycle in the vicinity of the time steps.} 
    \label{fig: finite size setup}
\end{figure}

We comment on the approach we take to extract the late time chiral mean flow per cycle in a finite size system. Given the geometry shown in Figure \ref{fig: finite size setup}, there will be both chiral transport and diffusive transport in our measurement protocol away from the perfect swapping case or away from the Zeno limit. In an infinitely large half-filled system, both chiral and diffusive transport will continue forever without significant boundary effect from the lattice. Nonetheless, in a finite size system, the late time flow of our measurement protocol can be altered after significant amount of particle diffuses to the boundary of the Lieb lattice rather than transported solely via chiral motion (see Fig \ref{fig: finite size setup}).

To account for this effect, we therefore extract the chiral flow rate of our lattice system by averaging about 5 cycles before the \textit{cumulative} density of particles at the upper right half edge (see Fig \ref{fig: finite size setup}) becomes significantly populated at some cutoff total density $\rho_{cutoff} \sim 0.1$. Here we track the \textit{cumulative} sum of density of particles that has \textit{ever} arrived at these sites as we immediately extract these particles after each protocol step. We then extract a $t_{cutoff}$ that happens when $\rho_{cutoff}$ reaches $0.1$ and we averaged 5 measurement cycles around $t_{cutoff}$ for our late time dynamics.

In the main article, except in the particular case of site vacancy disorder in the perfect swapping Zeno limit, where the protected chiral edge flow dynamics is deterministic ($p_{hop} = 1$) and not random (thereby diffusive dynamics is absent), we generally apply this approach for the extraction of the late time chiral flow dynamics. The diffusive behaviour is present in all other disorder cases (both in and out of Zeno limit).


\section{Derivation of the stochastic transition matrix for the random onsite potential disorder} \label{Appendix: Onsite}

We outline the derivation of Eq \ref{eq: disorder onsite stochastic} in this Appendix. We start off with the matrix form of the Hamiltonian and focusing on 2 unmeasured sites $\mathbf{r}$ and $\mathbf{r'}$ in the Zeno limit, where there is no other hopping elements coming into sites $\mathbf{r}$ and $\mathbf{r'}$
\begin{eqnarray}\small
    H_{zeno} =\left(
    \begin{array}{cccccccc}
     &  &  & 0 & 0 &  &  & \\
     &  &  & \vdots & \vdots &  &  & \\
     &  &  & 0 & 0 &  &  & \\
    0 & \ldots & 0 & W_{\mathbf{r}} & -t_{hop} & 0 & \ldots & 0 \\
    0 & \ldots & 0 & -t_{hop}  & W_{\mathbf{r'}} & 0 & \ldots & 0 \\
     &  &  & 0 & 0 &  &  & \\
     &  &  & \vdots & \vdots &  &  & \\
     &  &  & 0 & 0 &  &  & 
    \end{array}
    \right) .
\end{eqnarray}
Here, we note that the Zeno limit measurement effectively decoupled sites $\mathbf{r}$ and $\mathbf{r'}$ from the dynamics of the rest of the Hamiltonian. The time evolution unitary will retain the same decoupled form so for convenience, we only retain a $2 \times 2$ matrix for the rest of the derivation.

The resulting unitary matrix $U_{zeno} = \exp(- i H_{zeno} t)$ acting on the two sites when the Hamiltonian $H_{zeno}$ is allowed to evolve for time $t = \frac{T}{8}$ now takes the form (calculated using Mathematica)
\begin{eqnarray}\small
    U_{zeno} &=& \left(
    \begin{array}{cc}
    U_{11} & U_{12} \\
    U_{21} & U_{22}
    \end{array}
    \right) \text{ , } \\
    U_{11} &=& \exp( - i \frac{T}{8}  \frac{(W_{\mathbf{r}} + W_\mathbf{r'})}{2} ) \left( \cos(\frac{1}{2} \frac{T}{8} \sqrt{4 t_{hop}^2 + (W_{\mathbf{r}} - W_{\mathbf{r'}})^2 } ) - \frac{i \sin(\frac{1}{2} \frac{T}{8} \sqrt{4 t_{hop}^2 + (W_{\mathbf{r}} - W_{\mathbf{r'}})^2 }) (W_{\mathbf{r}} - W_{\mathbf{r'}})}{\sqrt{4 t_{hop}^2 + ( W_{\mathbf{r}} - W_{\mathbf{r'}} )^2}} \right) \nonumber \\
    U_{22} &=& \exp( - i \frac{T}{8}  \frac{(W_{\mathbf{r}} + W_{\mathbf{r'}})}{2} ) \left( \cos(\frac{1}{2} \frac{T}{8} \sqrt{4 t_{hop}^2 + (W_{\mathbf{r}} - W_{\mathbf{r'}})^2 } ) + \frac{i \sin(\frac{1}{2} \frac{T}{8} \sqrt{4 t_{hop}^2 + (W_{\mathbf{r}} - W_{\mathbf{r'}})^2 }) (W_{\mathbf{r}} - W_{\mathbf{r'}})}{\sqrt{4 t_{hop}^2 + ( W_{\mathbf{r}} - W_{\mathbf{r'}} )^2}} \right) \nonumber \\
    U_{12} = U_{21} &=& \frac{2 i \exp( - i \frac{T}{8}  \frac{(W_{\mathbf{r}} + W_{\mathbf{r'}})}{2} ) \sin(\frac{1}{2} \frac{T}{8} \sqrt{4 t_{hop}^2 + (W_{\mathbf{r}} - W_{\mathbf{r'}})^2 })  }{\sqrt{4 t_{hop}^2 + (W_{\mathbf{r}} - W_{\mathbf{r'}})^2}} \nonumber .
\end{eqnarray}

For an initial distribution $G = diag( g_1, g_2 )$, one then applies $G \rightarrow U G U^{\dagger}$ and a subsequent measurement to delete the resulting off-diagonal components in $G$. The resulting form of $G$ after the sequence of modified Zeno evolution and subsequent measurement takes the form
\begin{eqnarray}
    &\left(
    \begin{array}{cc}
    g_1 & 0 \\
    0 & g_2
    \end{array}
    \right) \rightarrow
    \left(
    \begin{array}{cc}
    g'_1 & 0 \\
    0 & g'_2
    \end{array}
    \right) \\
    &g'_1 = \frac{g_1 (2 t_{hop}^2 (1 + \cos(\frac{T}{8} \sqrt{4 t_{hop}^2 + ( W_{\mathbf{r}} - W_{\mathbf{r'}} )^2} ) ) + ( W_{\mathbf{r}} - W_{\mathbf{r'}} )^2 ) + g_2 ( 2 t_{hop}^2 (1 - \cos(\frac{T}{8} \sqrt{4 t_{hop}^2 + ( W_{\mathbf{r}} - W_{\mathbf{r'}} )^2} ) ) )}{4 t_{hop}^2 + (W_{\mathbf{r}} - W_{\mathbf{r'}} )^2} \nonumber \\  
    &g'_2 = \frac{g_2 (2 t_{hop}^2 (1 + \cos(\frac{T}{8} \sqrt{4 t_{hop}^2 + ( W_{\mathbf{r}} - W_{\mathbf{r'}} )^2} ) ) + ( W_{\mathbf{r}} - W_{\mathbf{r'}} )^2 ) + g_1 ( 2 t_{hop}^2 (1 - \cos(\frac{T}{8} \sqrt{4 t_{hop}^2 + ( W_{\mathbf{r}} - W_{\mathbf{r'}} )^2} ) ) )}{4 t_{hop}^2 + (W_{\mathbf{r}} - W_{\mathbf{r'}} )^2} \nonumber .
\end{eqnarray}
From these, one can readily extract the elements of the stochastic transfer matrices to take the form
\begin{eqnarray}
    &R_{i}=\oplus_{\langle \mathbf{r} , \mathbf{r'} \rangle\in  A_i} \left(
    \begin{array}{cc}\small
        1-p_{hop, \mathbf{r} \mathbf{r'}}  &  p_{hop, \mathbf{r} \mathbf{r'}}  \\
        p_{hop, \mathbf{r} \mathbf{r'}}  & 1-p_{hop, \mathbf{r} \mathbf{r'}} \\
    \end{array}
    \right) \oplus_{\text{other sites}} I , \nonumber \\ 
    &p_{hop, \mathbf{r} \mathbf{r'}} = \frac{2 t_{hop}^2 (1 - \cos( \frac{T}{8}\sqrt{4 t_{hop}^2 + (W_{\mathbf{r}} - W_{\mathbf{r'}} )^2 } ) )}{4 t_{hop}^2 + (W_{\mathbf{r}} - W_{\mathbf{r'}})^2}
\end{eqnarray}
which is Eq \ref{eq: disorder onsite stochastic} in the main text.

\end{document}